\documentclass[a4paper,12pt]{article}
\usepackage{amsmath,amssymb}
\usepackage{hyperref}

\usepackage{graphicx}
\usepackage{mathrsfs}
\usepackage{float}
\usepackage{ulem}
\usepackage{enumerate}
\usepackage{color}

\newcommand{\be}{\begin{equation}}
\newcommand{\ee}{\end{equation}}
\newcommand{\f}{\frac}
\newcommand{\s}{\sqrt}
\newcommand{\p}{\partial}

\newcommand{\bea}{\begin{equation}\begin{aligned}}
\newcommand{\eea}{\end{aligned}\end{equation}}
\newcommand{\ba}{\begin{align}}
\newcommand{\ea}{\end{align}}

\begin{document}

\begin{titlepage}

\vspace{.4cm}
\begin{center}
\noindent{\Large \textbf{Relative entropy of excited states in two dimensional conformal field theories}}\\
\vspace{1cm}
G\'abor S\'arosi$^{a,}$\footnote{sarosi@phy.bme.hu}
and
 Tomonori Ugajin$^{b,}$\footnote{ugajin@kitp.ucsb.edu}

\vspace{.5cm}
 {\it
 $^{a}$Department of Theoretical Physics, Institute of
Physics, Budapest University of Technology,  \\Budapest,
H-1521, Hungary\\
\vspace{0.2cm}
 }
\vspace{.5cm}
  {\it
 $^{b}$Kavli Institute for Theoretical Physics, University of California, \\
Santa Barbara, 
CA 93106, USA\\
\vspace{0.2cm}
 }
\end{center}


\begin{abstract}
We study the relative entropy and the trace square distance, both of which measure the distance between reduced density matrices of two excited states in two dimensional conformal field theories. We find a general formula 
for the relative entropy between two primary states with the same conformal dimension in the limit of a single small interval and find that in this case the relative entropy is proportional to the trace square distance.
We check our general formulae by calculating the relative entropy between two generalized free fields and the trace square distance between the spin and disorder operators of the critical Ising model. We also give the leading term of the relative entropy in the small interval expansion when the two operators have different conformal dimensions. This turns out to be universal when the CFT has no primaires lighter than the stress tensor. The result reproduces the previously known special cases.
\end{abstract}

\end{titlepage}

\tableofcontents

\vspace{1cm}

\newpage

\section{Introduction}

The AdS/CFT correspondence \cite{Maldacena:1997re, Witten:1998qj, Gubser:1998bc} can be regarded as a concrete formulation of 
a theory of quantum gravity in asymptotically anti-de Sitter space in terms of 
a conformal field theory living on the boundary. 

\vspace{0.1 cm}

This equivalence immediately leads to 
the following puzzle about the spectrum of these two theories. In the side of the large $N$ conformal field theory, there are lots of heavy states with fixed conformal dimension of order $N^2$. In two dimensions the density of these states is given by the Cardy formula. 
Yet, in the bulk gravity theory, it appears that there is only a single state, i.e. an AdS black hole with fixed energy. 
This observation leads us to the question of distinguishability of two black hole microstates. Indeed, this can be regarded as the foundation of the statistical interpretation of black hole entropy 
\cite{Strominger:1996sh,Callan:1996dv,Strominger:1997eq}, and it has been shown that correlation functions of these microstates cannot be distinguished from those of the thermal ensemble with the same energy 
\cite{Balasubramanian:2005qu, Balasubramanian:2005mg, Balasubramanian:2007qv, Das:2008ka}.

\vspace{0.1 cm}

This non-distinguishability property of black hole microstates plays an important role when we think about the
information loss problem \cite{Hawking:1976ra}. This is because one of the key assumptions in Hawking's calculation \cite{Hawking:1974sw} is that one can use  quantum field theory on a fixed black hole background to capture the evaporation process. From the point of view of the dual CFT, this means that one assumes that the difference between the microstates creating the background is always small during the evaporation. In principle, one can 
check whether this assumption holds or not by CFT calculations\footnote{One possible resolution of this non-distinguishability problem could be the fuzzball proposal \cite{Mathur:2005zp}, which states that in the bulk each black hole microstate could be significantly different at the horizon scale. 
 Here we would like to formulate the problem entirely within the CFT framework. }.

\vspace{0.1 cm}

In general, the CFT observers can access only a limited amount of information by measurements, forcing them to describe the system by (coarse grained) reduced density matrices. We will model this coarse graining by 
restricting our measurements to a particular region $A$ of the time slice of the CFT, i.e. our excited states will be described by reduced density matrices derived by tracing out the complement of $A$.

\vspace{0.1 cm}

With this background in mind, in this paper, we study two metric like structures which enable us to measure the distance between two reduced density matrices in two dimensional conformal field theories. By calculating these quantities we will see to what extent two reduced density matrices are distinguishable by measurements confined to region $A$. There are two quantities that we consider for this purpose.

\vspace{0.1 cm}

One measure is the \textit{relative entropy} $S(\rho|| \sigma)$ between two reduced density matrices $ \rho$ and $ \sigma$, which is defined by 
\be
S(\rho|| \sigma)=   {\rm tr}\; \rho\log \rho-{\rm tr}\; \rho\log \sigma.
\ee
Note that this quantity is automatically free from UV divergences. Another intriguing property of the relative entropy is its positive definiteness. This property has been efficiently used to shed light on some aspects of theories coupled to semiclassical gravity, including the precise formulation of the Bekenstein bound \cite{Casini:2008cr}, proof of the generalized second law\cite{Wall:2010cj,Wall:2011hj}, the quantum Bousso bound \cite{Bousso:2014sda,Bousso:2014uxa}, and when applied to theories with a CFT dual, this positivity 
is related to certain positivity conditions of the bulk stress tensor \cite{Lin:2014hva,Lashkari:2014kda,Lashkari:2015hha}.
\noindent
There are also related works on relative entropy involving holography \cite{Blanco:2013joa, Lashkari:2013koa, Faulkner:2013ica, Jafferis:2015del}. 


With the aim of a purely CFT discussion, in a recent paper \cite{Lashkari:2015dia} a replica trick to compute this quantity efficiently was discussed. Using this trick, the calculation of $S(\rho|| \sigma)$ boils down to the computation of a bunch of correlation 
functions on a Riemann surface with cuts. This formalism was used to study the relative entropy between 
two excited states in the free boson theory in two dimensions \cite{Lashkari:2015dia}. 
See also \cite{Lashkari:2014yva} for an earlier discussion.  The computation is very similar to the computation of entanglement entropy of excited states, see for example \cite{Bhattacharya:2012mi, Alcaraz:2011tn,Caputa:2014vaa, Asplund:2014coa, Nozaki:2014uaa,Caputa:2014eta,Caputa:2015waa,Mosaffa:2012mz,Giusto:2014aba}.

\vspace{0.1 cm}
 
The other measure that we discuss in this paper is what we call \textit{trace square distance}
\be
T(\rho||\sigma)=\f{{\rm tr}\; |\rho-\sigma|^2}{{\rm tr}\; \rho_{(0)}^2}, 
\ee
where ${\rm tr}\; \rho_{(0)}^2 $ denotes the second R\'enyi entropy of the vacuum. This factor removes the unwanted UV divergence in the numerator coming from short distance entanglement. Notice that this is not the usual trace distance which is frequently used in the quantum information community \cite{NC}. 
Nevertheless, we will find that this quantity is still useful. This is partially because it is computable once we know certain four point functions on the two sheeted replica manifold and hence higher $n$ replicas or analytic continuations are not needed. Furthermore, in certain cases that we will consider below, it essentially captures the behavior of the relative entropy in the small subsystem limit. 

\vspace{0.1 cm}
The main statements of this paper are the following.  Let  $|V \rangle$, $|W \rangle $ be primary excited states of a CFT on a cylinder and $A$ be the subsytem defined on a time slice of the cylinder. One can consider reduced density matrices associated with the excited states on the region $A$, by tracing out the complement of $A$ with respect to the time slice
\be
\rho_{V}= {\rm tr}|_{A_{c}} |V \rangle \langle V| \quad \rho_{W}= {\rm tr}|_{A_{c}} |W \rangle \langle W|.
\ee
Also, when the cylinder is mapped to the plane, one can identify these states with some primary operators using the 
state operator correspondence.
Then,

\vspace{0.1 cm}
\begin{enumerate}[(i)]
\item For any 2d CFT, and any pair of primary states $|V \rangle$ and $|W \rangle$ with the same conformal dimension\footnote{In this paper we consider excited states without spin, $h=\bar{h}$. The generalization to $h \neq \bar{h}$ should be trivial.} $h_{V}=h_{W}$, the relative entropy in the small interval limit 
$|A| =2\pi x \ll 1$ is given by 
\be
\label{eq:relent1}
S(\rho_{V}||\rho_{W})=\frac{\Gamma(\f{3}{2})\Gamma(\Delta+1)}{2\Gamma(\Delta+\f{3}{2})}\sum_{\alpha}\left(C_{O_{\alpha}VV}-C_{O_{\alpha}WW} \right)^2 (\pi x)^{2\Delta}+\cdots, 
\ee
where $\{O_{\alpha} \}$ is the set of the lightest primary operators with $C_{O_{\alpha}VV}-C_{O_{\alpha}WW} \neq 0$, $\Delta=h_\alpha+\bar h_\alpha$ is the scaling dimension and $\cdots$ stands for powers of $x$ larger than $2\Delta$. We choose the operators 
in this set to satisfy the orthonormality condition $\langle O^{\dagger}_{\alpha}(\infty) O_{\beta}(0) \rangle  =\delta_{\alpha \beta}$, without the loss of generality.

\vspace{0.1 cm}

\item In the same limit with $h_{V}=h_{W}$, the relative entropy is related to the trace square distance $T(\rho_{V}||\rho_{W})$ by
\be
S(\rho_{V}||\rho_{W})=2^{2\Delta -1} \frac{\Gamma(\f{3}{2})\Gamma(\Delta+1)}{\Gamma(\Delta+\f{3}{2})} \;T(\rho_{V}||\rho_{W}).
\ee
\vspace{0.1 cm}

\item \label{point3} Formula \eqref{eq:relent1} also applies when the conformal dimensions are different $h_V\neq h_W$ provided that the lightest primaries $\{O_{\alpha} \}$ with $C_{O_{\alpha}VV}-C_{O_{\alpha}WW} \neq 0$ have scaling dimensions $\Delta=h_\alpha +\bar h_\alpha<2$, i.e. they are lighter than the stress tensor, or equivalently, they are relevant.

\vspace{0.1 cm}

\item In the case when all the primaries with $C_{O_{\alpha}VV}-C_{O_{\alpha}WW} \neq 0$ have scaling weight $\Delta>2$ (e.g. there is a gap in the primary spectrum), the relative entropy of any pair of primary states $|V \rangle$ and $|W \rangle$ with $h_V\neq h_W$ in the small interval limit $|A| =2\pi x \ll 1$ takes the following universal form
\be
S(\rho_{V}||\rho_{W})=\f{16}{15}\f{1}{c}\left(h_V-h_W \right)^2 (\pi x)^{4} +\cdots,
\ee
where $\cdots$ stands for powers of $x$ larger than $4$. 
\end{enumerate}

\vspace{0.1 cm}

These results are all consistent with  previously known results \cite{Blanco:2013joa,Lashkari:2015dia,Lashkari:2014yva,Asplund:2014coa}.
We also check them by computing the relative entropy between two generalized free fields directly, without using the above formulae. The generalized free fields are low energy excitations of CFTs with a gravity dual in the large central charge limit. In the gravity side, they are identified with free scalar fields in the bulk. Since the bulk theory is free, one can compute their correlation functions by using Wick contractions.

\vspace{0.1 cm}

This paper is organized as follows. In section \ref{sec:def} we explain how to compute the relative entropy $S(\rho||\sigma) $ and the trace square 
distance $T(\rho|| \sigma)$ by using the replica trick of \cite{Lashkari:2015dia}. In section \ref{sec:generaltheory} we compute these quantities in a general CFT in the small interval limit and prove the above statements. 
In section 
\ref{sec:generalizedfreefields} we calculate the relative entropy and the trace square distance for generalized free fields and in section \ref{sec:ising} we calculate the trace square distance between the spin and disorder fields of the critical two dimensional Ising model. We use these results to check our fomulae. In section \ref{sec:conc} we comment on possible applications of the results and 
conclude the paper.

\section{Review of the basic definitions}\label{sec:def}

There has been a considerable amount of interest in studying entanglement properties of quantum field theories in recent years. Particularly powerful techniques have been developed with this aim in conformal field theories. Among such techniques are realizations of the replica trick via the uniformization map or via the cyclic orbifold CFT, which have been efficiently used to calculate e.g. the R\'enyi and entanglement entropy\cite{Holzhey:1994we,Calabrese:2004eu,Cardy:2007mb, Calabrese:2009qy,Berganza:2011mh,Alcaraz:2011tn,Calabrese:2009ez, Calabrese:2010he} or the entanglement negativity\cite{Calabrese:2012ew,Calabrese:2014yza} of various states on various subregions. Here, we briefly review these techniques and highlight some key ingedients to keep in mind when one wishes to apply them to calculate distance measures, like the relative entropy. 

\subsection{Reduced density matrices of excited states}

\begin{figure}[h!]
\centering
\includegraphics[width=1\textwidth]{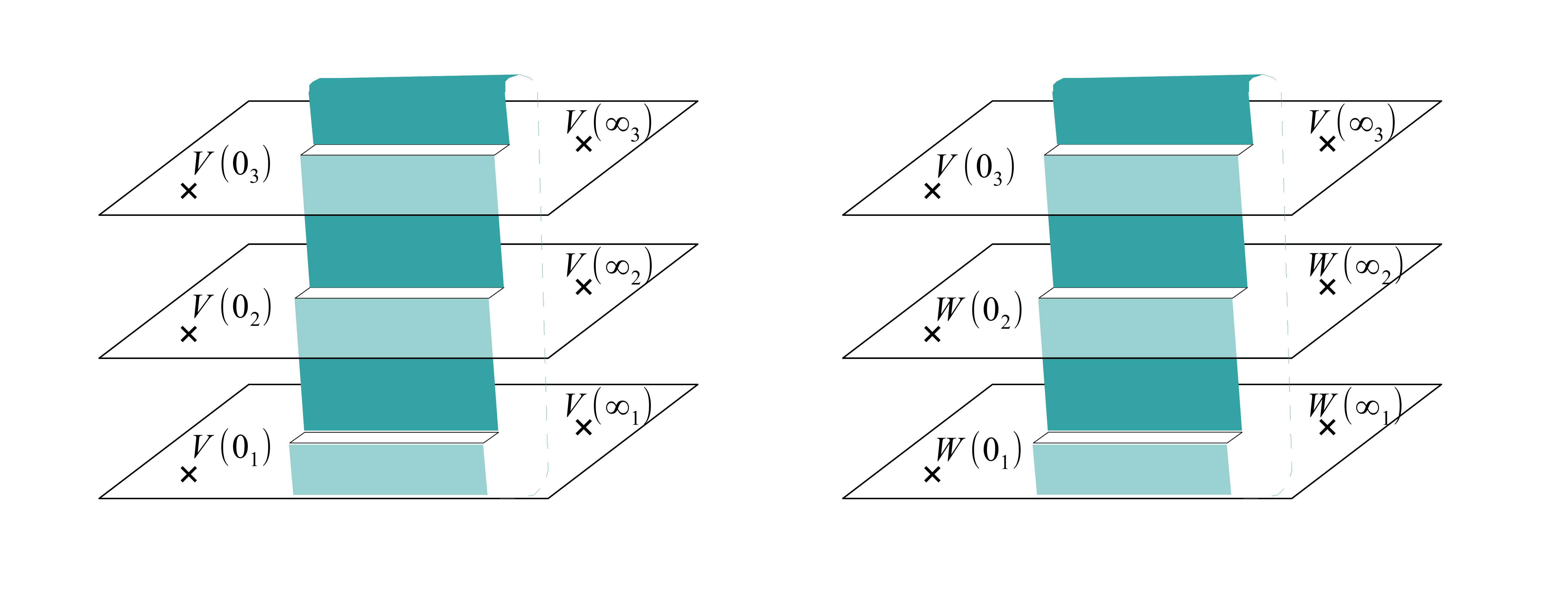}
\caption{Illustration of the relevant correlation functions for the 3-sheeted manifold $\Sigma_3$. The left figure corresponds to $\text{Tr}\rho_V^3$ while the right one to $\text{Tr}\rho_V \rho_W^2$. The cyclic $Z_3$ replica symmetry becomes the cyclic symmetry of the trace.}
\label{fig:1}
\end{figure}

Consider a CFT on a cylinder. Let $(T, \phi) \sim (T,\phi+2\pi) $ be the (Euclidean) timelike and the spacelike coordinate of the cylinder respectively. 
The subsystem $A$ is defined to be the segment $[0, 2\pi x]$ at $T=0 $. Let us denote the reduced density matrix of a state $| V \rangle $ to the subsystem $A$ by $\rho_V$. The $n$-th R\'enyi entropy of this density matrix, $\text{Tr}\rho_V^n$ is given by the transition amplitude of $n$ copies of $ | V \rangle$ between   $T=-\infty$
and $T= \infty$, on the $n$ sheeted cover of the cylinder glued together along the cut corresponding to the subsystem $A$. Equivalently, one can regard the system to be defined on an $n$ sheeted covering of the plane $\Sigma_{n}$ using the exponential map $z= e^{T+i\phi}$.  On this $n$ sheeted plane, the $n$ copies of the excited state are located at the origin and infinity of the each sheet. Because of the state operator 
correspondence, we can create the state on the each sheet by a local operator
\be
 | V \rangle=V(0)|0 \rangle, \quad \langle V |= \langle 0| V^{\dagger} (\infty)= \lim_{z \rightarrow \infty}\langle 0| V (z) z^{2\bar{h}_{V}} \bar{z}^{2h_{V}},
\ee
where $V^{\dagger}$ denotes the BPZ conjugate.
The cut on the $n$ sheeted cylinder is mapped to a segment on the unit radius circle of the $n$ sheeted plane $\Sigma_{n}$ whose end points are
\be
u=e^{2\pi ix}, \quad v=1.
\ee
Then, the trace of the reduce density matrix ${\rm tr}\;\rho^{n}_{V}$ can be written as a $2n$ point  function 
 on $\Sigma_{n}$ 
\be
{\rm tr}\; \rho^{n}_{V}= {\rm tr}\; \rho^{n}_{(0)} \;\langle \prod_{k=0}^{n-1} V^{\dagger}(\infty_{k}) \prod_{k=0}^{n-1}V(0_{k})\rangle_{\Sigma_{n}}, \label{eq:reddenV}
\ee
where ${\rm tr}\; \rho^{n}_{(0)} $ denotes the partition function on $\Sigma_{n}$. The operator locations $\{ \infty_{k}, 0_{k} \}$ denote the infinity and the origin of the $k$-th sheet respectively, see Fig. \ref{fig:1}. 

There are two ways to study this $2n$ point function. One way is to use the equivalence between a CFT $\mathcal{C}$ on $\Sigma_{n}$ and its cyclic orbifold theory $\mathcal{C}^n/Z_{n}$ on the plane.

\noindent We can write this $2n$ point function on $\Sigma_{n}$ as a four point function involving the twist operators  $\sigma_{n}$,
\be
\langle \prod_{k=0}^{n-1} V^{\dagger}(\infty_{k}) \prod_{k=0}^{n-1}V(0_{k})\rangle_{\Sigma_{n}} =\frac{\langle V^{\dagger}_{n}(\infty) \sigma_{n}(1) \tilde{\sigma}_{-n}(\tilde{w}) V_{n}(0)\rangle_{\mathcal{C}^n/Z_{n}}}{\langle \sigma_{n}(1) \tilde{\sigma}_{-n}(\tilde{w}) \rangle}, \label{eq:4ptvn}
\ee
where
\be
V_{n}:=V^{\otimes n}, \quad \tilde{w}=\f{u}{v} =e^{2\pi i x},
\ee
and the conformal dimension of the twist operator is given by $h_{\sigma_{n}}= \bar{h}_{\sigma_{n}}= cn(1-1/n^2)$, where $c$ is the central charge of the seed theory $\mathcal{C}$.

The other way to study this $2n$ point function is to use the uniformalization map between the plane and its $n$ sheet cover $\Sigma_{n}$ \footnote{We denote the holomorphic coordinates on the plane and its $n$ sheet cover $\Sigma_{n}$ by $w$ and $z$ respectively.}. This map is given by 
\be
\label{eq:unifmap}
w(z)=\sqrt[n]{\f{z-u}{z-v}}.
\ee
According to this map each operator on $\Sigma_{n}$ is mapped to the location
\be
\infty_{k} \rightarrow w_{k}=e^{\f{2\pi i k}{n}}, \quad 0_{k} \rightarrow \hat{w}_{k} =e^{\f{2\pi i (k+x)}{n}}.
\ee
on the uniformized plane.
By using the Jacobian factor
\begin{align}
D(k, h_{V}) = \lim_{z_{k} \to \infty} |z_{k}|^{4h_{V}}\Big| \f{\p w}{\p z} \Big|^{2h_{V}}_{\infty_{k}} \Big| \f{\p w}{\p z} \Big|^{2h_{V}}_{0_{k}} =\left[ \f{2}{n}\sin \pi x \right]^{4h_{V}}, \label{eq: Jacobian}
\end{align}
we find the relation between the $2n$ point function (\ref{eq:reddenV}) on $\Sigma_{n}$ and
the $2n$ point function on the plane
\be
\label{eq:uniformcorrelator}
 \langle \prod_{k=0}^{n-1} V^{\dagger}(\infty_{k}) \prod_{k=0}^{n-1}V(0_{k})\rangle_{\Sigma_{n}} =\left[ \f{2}{n}\sin \pi x\right]^{4nh_{V}}\langle \prod_{k=0}^{n-1} V(w_{k}) \prod_{k=0}^{n-1}V(\hat{w}_{k})\rangle.
\ee
Notice that in the small subsystem limit $x \rightarrow 0$, each $w_{k}$ approaches $\hat{w_{k}}$, therefore the $2n$ point  
function is factorized into a product of two point functions
\be
\label{eq:factorization}
\langle \prod_{k=0}^{n-1} V(w_{k}) \prod_{k=0}^{n-1}V(\hat{w}_{k})\rangle=\prod_{k=0}^{n-1} 
\langle V(w_{k})V(\hat{w}_{k})\rangle.
\ee

\subsection{Relative entropy}

Now let us consider the relative entropy between two reduced density matrices $\rho, \sigma$
\be
S( \rho|| \sigma)=  {\rm tr}\; \rho\log \rho-{\rm tr}\; \rho\log \sigma.
\ee
A replica trick to compute this was introduced in \cite{Lashkari:2015dia}

\begin{align} 
S( \rho|| \sigma) &=\lim_{n \rightarrow 1} S_{n}( \rho|| \sigma) \nonumber \\
&= \lim_{n \rightarrow 1} \f{1}{n-1} \left(\log {\rm tr}\; \rho^n -\log  {\rm tr}\;\rho \sigma^{n-1} \right) \label{eq:relplica}.
\end{align} 
We may use this to express the relative entropy between the reduced density matrices $\rho_{V}, \rho_{W}$ of two excited 
states
\bea
|V \rangle &=V(0)|0 \rangle, && |W \rangle &=W(0)|0 \rangle.
\eea
with CFT correlation functions. We can write the trace involving different RDMs as
\begin{align}
 {\rm tr}\;\rho_{V} \rho_{W}^{n-1} &=  {\rm tr}\; \rho^{n}_{(0)} \; \langle V^{\dagger}(\infty_{0}) \prod_{k=1}^{n-1} W^{\dagger}(\infty_{k}) V(0_{0}) \prod_{k=1}^{n-1}W(0_{k})\rangle_{\Sigma_{n}}, \nonumber \\[+8pt]
&=\langle X^{\dagger}_{n}(\infty) \sigma_{n}(1) \tilde{\sigma}_{-n}(\tilde{w}) X_{n}(0)\rangle_{\mathcal{C}^n/Z_{n}},
\quad X_{n}=V \otimes W^{\otimes(n-1)}. \label{eq:4ptxn}
\end{align}

\vspace{0.3cm}
\noindent
Notice that the operator $X_n$ is not symmetrized under cyclic permutations and hence it is not part of the spectrum of the orbifold theory $\mathcal{C}^n/Z_{n}$. 
Nevertheless, it is still useful to introduce an expression for (\ref{eq:4ptxn}) involving the twist operators, like (\ref{eq:4ptvn}), as we will now explain. To do this, it is convenient to first transform $\Sigma_{n}$ by the global transformation, 
\be
y(z)=\f{z-\tilde{w}}{z-1}.
\ee
Now the cut is extending between $0$ and $\infty$. Each operator on the $k$th sheet is mapped to
\be
\infty_{k} \rightarrow 1_{k}, \quad 0_{k} \rightarrow \tilde{w}_{k},
\ee
where $\{ 1_{k}, \tilde{w}_{k} \}$ denote $1$ and $e^{2 \pi i x}$ on the $k$th sheet respectively. Using this, we can write the operator product expansion of the operator insertions on each sheet as
\begin{align}
V^{\dagger} (\infty_{k}) V(0_{k}) &= \lim_{z_{k} \rightarrow \infty} |z_{k}|^{4h_{V}} \Big| \f{\p y}{\p z} \Big|_{z_{k}}^{2h_{V}} \Big| \f{\p y}{\p z}\Big|_{0_{k}}^{2h_{V}} V(1_{k}) V(\tilde{w}_{k}) \nonumber \\[+10pt]
&= \sum_{T_{k}}C^{T_{k}}_{VV}  (1-\tilde{w})^{h_{T_{k}}}  (1-\bar{\tilde{w}})^{\bar{h}_{T_{k}}} T_{k}(1_{k}) , \label{eq:OPE}
\end{align}
where $T_{k}$ denotes a state in the seed theory. 
A similar expansion holds for the $W^{\dagger} \times W$ OPE.  We substitute these OPEs into the first line of (\ref{eq:4ptxn}) which yields a sum over correlators on $\Sigma_n$ with only a single insertion on each sheet. Applying the cyclic orbifold prescription to each of these correlators and grouping of the terms results in the formula 
\begin{align}
 {\rm tr}\;\rho_{V} \rho_{W}^{n-1} &= \sum_{T} \hat{C}_{X_{n}X_{n}}^{T} \hat{C}_{T \sigma_{n} \tilde{\sigma}_{-n}}  (1-\tilde{w})^{h_{T}} (1-\bar{\tilde{w}})^{\bar{h}_{T}}.
\end{align}
Here, $T$ denotes an \textit{untwisted} state in the orbifold theory, 
\be
T= \left[ T_{1} \otimes \cdots \otimes T_{n} \right]_{{\rm sym}} ,
\ee
where  $[\cdots]_{{\rm sym}}$ means that we symmetrize the operator in the parenthesis under cyclic permutations by summing up all orbits and divide it by $n$. 
The coefficients $\hat{C}_{X_{n}X_{n}}^{T}, \hat{C}_{T \sigma_{n} \tilde{\sigma}_{-n}}$ are defined by
\be
\hat{C}_{X_{n}X_{n}}^{T} \equiv n \left[ C_{VV}^{T_{1}} \prod_{k=2}^{n} C^{T_{k}}_{WW}  \right]_{{\rm sym}} \quad 
\hat{C}_{T \sigma_{n} \tilde{\sigma}_{-n}} = \langle \sigma_{n}( \infty) T(1) \tilde{\sigma}_{-n}(0) \rangle_{\mathcal{C}^n/Z_{n}}. \label{eq:3pfn}
\ee
Based on this, we can \textit{define} the four point function in the second line of (\ref{eq:4ptxn}) as
\begin{align}
\langle X^{\dagger}_{n}(\infty) \sigma_{n}(1) \tilde{\sigma}_{-n}(\tilde{w}) X_{n}(0)\rangle_{\mathcal{C}^n/Z_{n}} &\equiv  \sum_{T} \hat{C}_{X_{n}X_{n}}^{T} \hat{C}_{T \sigma_{n} \tilde{\sigma}_{-n}}  (1-\tilde{w})^{h_{T}-2h_{\sigma_{n}}} (1-\bar{\tilde{w}})^{\bar{h}_{T}-2h_{\sigma_{n}}}, \nonumber \\
 X_{n}& \equiv V \otimes W^{\otimes(n-1)}. \label{eq:def4pt}
\end{align}
Although we will use this four point function because of the notational simplicity, we should keep in mind that $X_{n}$ is not an operator in the orbifold theory, therefore the definition is rather formal. Whenever the left hand side of (\ref{eq:def4pt}) appears, this always means the right hand side of (\ref{eq:def4pt}).

We can write $S_{n}( \rho_{V}|| \rho_{W})$ in terms of the correlation functions of $V$ and $W$ in the following three ways

\vspace{0.3cm}

\begin{align}
S_{n}( \rho_{V}|| \rho_{W})&= \f{1}{n-1}\log \frac{\langle \prod_{k=0}^{n-1} V^{\dagger}(\infty_{k}) \prod_{k=0}^{n-1}V(0_{k})\rangle_{\Sigma_{n}} }{ \langle V^{\dagger}(\infty_{0}) \prod_{k=1}^{n-1} W^{\dagger}(\infty_{k}) V(0_{0}) \prod_{k=1}^{n-1}W(0_{k})\rangle_{\Sigma_{n}}} \nonumber \\[+16pt]
&=\f{1}{n-1} \log \f{\langle V^{\dagger}_{n}(\infty) \sigma_{n}(1) \tilde{\sigma}_{-n}(\tilde{w}) V_{n}(0)\rangle_{\mathcal{C}^n/Z_{n}}}{\langle X^{\dagger}_{n}(\infty) \sigma_{n}(1) \tilde{\sigma}_{-n}(\tilde{w}) X_{n}(0)\rangle_{\mathcal{C}^n/Z_{n}}},  \nonumber \\[+16pt]
&= \f{1}{n-1}\log \left( \frac{\langle \prod_{k=0}^{n-1} V(w_{k}) \prod_{k=0}^{n-1}V(\hat{w}_{k})\rangle }{ \langle V(w_{0}) \prod_{k=1}^{n-1} W(w_{k}) V(\hat{w}_{0})\prod_{k=1}^{n-1}W(\hat{w}_{k})\rangle}\left[ \f{2}{n}\sin \pi x\right]^{4(n-1)(h_{V}-h_{W})}\right).
\label{eq:relentdef}
\end{align}

\subsection{Trace square distance}

Now consider the quantity
\be 
T(\rho_{V}|| \rho_{W})= \f{{\rm tr}\; |\rho_{V}-\rho_{W}|^2}{{\rm tr}\; \rho_{(0)}^2}.
\ee
It is clear that $T(\rho_{V}|| \rho_{W})$ is manifestly positive and zero only when $\rho_{V}=\rho_{W}$. We will call $T(\rho_{V}|| \rho_{W})$ the trace square distance in order to distinguish it from the trace distance, which in the quantum information literature is usually used for the distance in the nuclear norm $\text{Tr}\sqrt{AA^\dagger}$.
One can express the trace square distance with the use of four point functions as
\begin{align}
T(\rho_{V}|| \rho_{W})&= \langle V^{\dagger} (\infty_{1})  V^{\dagger} (\infty_{2})  V(0_{1})V(0_{2}) \rangle_{\Sigma_{2}}+\langle W^{\dagger} (\infty_{1})W^{\dagger} (\infty_{2}) W(0_{1}) W(0_{2}) \rangle_{\Sigma_{2}} \nonumber \\[8pt]
&-2\langle V^{\dagger} (\infty_{1})  W^{\dagger} (\infty_{2})  V(0_{1})W(0_{2}) \rangle_{\Sigma_{2}} \nonumber \\[8pt]
&=\f{\langle V^{\dagger\otimes 2}(\infty) \sigma_{2}(1) \tilde{\sigma}_{-2}(\tilde{w}) V^{\otimes 2}(0)\rangle_{\mathcal{C}^2/Z_{2}}}{\langle \sigma_{2}(1) \tilde{\sigma}_{-2}(\tilde{w})\rangle_{\mathcal{C}^2/Z_{2}}}+\f{\langle W^{\dagger\otimes 2}(\infty) \sigma_{2}(1) \tilde{\sigma}_{-2}(\tilde{w}) W^{\otimes 2}(0)\rangle_{\mathcal{C}^2/Z_{2}}}{\langle \sigma_{2}(1) \tilde{\sigma}_{-2}(\tilde{w})\rangle_{\mathcal{C}^2/Z_{2}}} \nonumber \\[8pt]
&-2\f{\langle V^{\dagger}\otimes W^{\dagger} (\infty)  \sigma_{2}(1) \tilde{\sigma}_{-2}(\tilde{w}) V\otimes W (0)\rangle_{\mathcal{C}^2/Z_{2}}}{\langle \sigma_{2}(1) \tilde{\sigma}_{-2}(\tilde{w})\rangle_{\mathcal{C}^2/Z_{2}}}, \label{eq:tsd}
\end{align}
where the four point function involving $V\otimes W$ is defined by \eqref{eq:def4pt}.

\section{$S(\rho_{V}|| \rho_{W})$ and $T(\rho_{V}|| \rho_{W})$ in the small interval limit} \label{sec:generaltheory}

In this section we would like to compute both the relative entropy $S(\rho_{V}|| \rho_{W})$ and the trace square distance
 $T(\rho_{V}|| \rho_{W})$ for generic two dimensional conformal field theories when the subsystem size $|A|=2\pi x$
is small.

\subsection{Contributions of the vacuum} 

In the limit $x<<1$ we expect the contribution of the vacuum exchange\footnote{In this subsection, by vacuum exchange, we mean the vacuum of the original theory and not the orbifold.} to dominate in each relevant correlation function which leads to factorization into two point funtions, see \eqref{eq:factorization}. Now we would like to identify this contribution. We will later see that it is possible that the leading behaviour is not determined by the vacuum contribution.

Let us consider the trace square distance first. It is decomposed into a sum over all states in the CFT,
\be
T(\rho_{V}|| \rho_{W}) =\left( \langle V^{\dagger}(\infty)V(0) \rangle_{\Sigma_{2}} -\langle W^{\dagger}(\infty)W(0)  \rangle_{\Sigma_{2}} \right)^2 + \cdots,
\ee
where $\cdots$ denotes the contribution of excited states.
Since the two point function on $\Sigma_n$ is given by
\be
\langle V^{\dagger}(\infty)V(0) \rangle_{\Sigma_{n}} = \left[ \f{\sin(\pi x)}{n\sin(\f{\pi x}{n})} \right]^{4h_{V}},
\ee
the contribution of the vacuum exchange to the trace square distance is
\be
T_{{\rm vac}} (\rho_{V}|| \rho_{W}) =\left[ \left(\cos \f{\pi x}{2} \right)^{4h_{V}} -\left(\cos \f{\pi x}{2} \right)^{4h_{W}}\right]^2.
\label{eq:tracedistfactor}
\ee
This expression can only be valid in the small subsystem limit $x<<1$ where it takes the form
\be
\label{eq:tracedistsmalldistvac}
T_{{\rm vac}}(\rho_{V}|| \rho_{W}) =\f{1}{4}(h_V-h_W)^2 (\pi x)^4+O(x^6).
\ee
It is important to note that although in the small subsystem limit the vacuum exchange dominates in each four point function in (\ref{eq:tsd}), it might happen that $T_{{\rm vac}} (\rho_{V}|| \rho_{W})$ does not give the leading term in the trace square distance in this limit. This is because the vacuum exchange contribution can cancel among the four point functions in (\ref{eq:tsd}) which manifestly happens when $h_V=h_W$.

This cancellation of the vacuum exchange contribution is {\it exact} in the case of the relative entropy: 
\be
S_{n}( \rho_{V}|| \rho_{W})  = \left[ \log \langle V^{\dagger}(\infty)V(0) \rangle_{\Sigma_{n}} -\log\langle W^{\dagger}(\infty)W(0) \rangle_{\Sigma_{n}} \right]+ \cdots.
\ee
In the $n \rightarrow 1$ limit this clearly vanishes.

\subsection{Small subsystem expansion} 

It is possible to give an expansion for $S_{n}( \rho_{V}|| \rho_{W})$ and $T( \rho_{V}|| \rho_{W})$ in the subsystem size $x$ in terms of the states of the orbifold theory $\mathcal{C}^n/Z_n$. Recall the orbifold prescription for the correlation functions (\ref{eq:4ptvn}) and (\ref{eq:4ptxn})
\be
G_{V_{n}}=\f{\langle V^{\dagger}_{n}(\infty) \sigma_{n}(1) \tilde{\sigma}_{-n}(\tilde{w}) V_{n}(0)\rangle_{\mathcal{C}^n/Z_{n}}}{ \langle \sigma_{n}(1) \tilde{\sigma}_{-n}(\tilde{w}) \rangle}, \quad G_{X_{n}}=\f{\langle X^{\dagger}_{n}(\infty) \sigma_{n}(1) \tilde{\sigma}_{-n}(\tilde{w}) X_{n}(0)\rangle_{\mathcal{C}^n/Z_{n}}}{ \langle \sigma_{n}(1) \tilde{\sigma}_{-n}(\tilde{w}) \rangle} .
\ee
These four point functions can be written, using the definition \eqref{eq:def4pt}, as a sum over all states in the orbifold theory $\mathcal{C}^n/Z_{n}$
\be
G_{X_{n}}= \langle X^{\dagger}_{n} (\infty) X_{n}(0) \rangle+\sum_{T \neq{\rm vac}} \hat{C}_{X_{n} X_{n}T} \hat{C}^{T}_{\sigma_{n}\tilde{\sigma}_{-n}} (1-\tilde{w})^{h_{T}} (1-\bar{\tilde{w}})^{\bar{h}_{T}}.
\ee
We can expand $G_{V_{n}}$ in a similar way.
In the small subsystem limit $1-\tilde{w} \sim 2\pi ix \ll 1$, the second term is much smaller than the first term and we can expand the logaritm
so 
\begin{align}
S_{n}( \rho_{V}|| \rho_{W}) &=\f{1}{n-1} \left(\log G_{V_{n}}- \log G_{X_{n}} \right) \nonumber \\
&\approx \f{1}{n-1}\sum_{T \neq{\rm vac}}(\hat{C}_{X_{n}X_{n}T}-\hat{C}_{V_{n} V_{n}T}) \hat{C}^{T}_{\sigma_{n}\tilde{\sigma}_{-n}} (1-\tilde{w})^{h_{T}} (1-\bar{\tilde{w}})^{\bar{h}_{T}}, \label{eq:replicarelative}
\end{align}
where the second line gives the correct leading behaviour in $x$.
Similarly, the trace square distance can be expanded as
\begin{align}
T( \rho_{V}|| \rho_{W}) &=G_{V_2}+G_{W_2}-2G_{X_2}\nonumber \\
&=\sum_{T \neq{\rm vac}}(\hat{C}_{V_{2} V_{2}T}+\hat{C}_{W_{2} W_{2}T}-2\hat{C}_{X_{2}X_{2}T}) \hat{C}^{T}_{\sigma_{2}\tilde{\sigma}_{-2}} (1-\tilde{w})^{h_{T}} (1-\bar{\tilde{w}})^{\bar{h}_{T}}. \label{eq:replicatracesquare}
\end{align}
Note that these sums are over all non-twisted sector states (not just primaries!) of the orbifold theory except its vacuum.

\subsection{Small interval limit for operators with equal weight}
\label{subsec:equalweight}

Now we would like to apply the formula \eqref{eq:replicarelative} to the case when the two operators have equal weight $h_V=h_W$. Let us first focus on the contribution of the states of the form
\be
T= \left[ \bigotimes_{k=0}^{n-1} L^{(k)}_{-\{m_{i_{k}}\}} \right]_{\rm sym},
\ee
where, $\{m_{i_{k}}\}=\{m_{1_{k}}, m_{2_{k}} , \cdots \}$ is a sequence of non negative integers,  $L^{(k)}_{-\{m_{k}\}}$ denotes a product of the Virasoro generators $L^{(k)}$
\be
\label{eq:virasoros}
L^{(k)}_{-\{m_{i_{k}}\}}=L^{(k)}_{-m_{1_{k}}}L^{(k)}_{-m_{2_{k}}} \cdots L^{(k)}_{-m_{l_{k}}} \cdots.
\ee
acting on $k$-th component of the tensor product. 
The OPE coefficients $\hat{C}_{X_{n} X_{n}T},\hat{C}_{V_{n} V_{n}T}$ of these states can be computed from the two point functions $ \langle X^{\dagger}_{n} (\infty) X_{n}(0) \rangle$, $\langle V^{\dagger}_{n} (\infty) V_{n}(0) \rangle$
by using the conformal Ward identity. This means that these OPE coefficients only depend on the conformal dimensions $h_{V},h_{W}$, therefore
\be
\hat{C}_{X_{n}X_{n}T}=\hat{C}_{V_{n}V_{n}T}, \quad {\rm when \;\;} h_{V}=h_{W}.
\ee
This shows that none of the vacuum descendants\footnote{Note that this argument also elliminates contribution from any primary of the orbifold theory which is built out of vacuum descendants of the original theory.} contribute to either (\ref{eq:replicarelative}) or \eqref{eq:replicatracesquare} when $ h_{V}=h_{W}$.
Therefore, the lightest states in the orbifold theory $\mathcal{C}^{n}/Z_{n}$  which can appear in the final result come from the lightest (non-vacuum) 
primary operators $\{ O_{\alpha} \}$ in the seed theory $\mathcal{C}$. Let us denote with $h_\alpha$ and $\bar h_\alpha$ the conformal weights, with $\Delta=h_\alpha+\bar h_\alpha$ the scaling dimension (independent of $\alpha$) and with $s_\alpha=h_\alpha-\bar h_\alpha$ the spin of these primary operators. Then, the lightest contributing orbifold operators have the following form\footnote{Note that operators of the form $\left[ O_{\alpha} \otimes I^{\otimes (n-1) }\right]_{\rm sym}$ do not contribute as their OPE coefficients with the twist fields are proportional to their one point functions on the plane which, of course, vanishes.} 
\be
\label{eq:biop}
O^{j}_{\alpha}= \left[ O_{\alpha} \otimes I^{\otimes (j-1)} \otimes O_{\alpha} \otimes I^{\otimes (n-j-1)} \right]_{{\rm sym}}, \quad j\leq \f{n}{2}
 \ee
The coefficients $\hat{C}_{O^{j}_{\alpha} X_{n} X_{n}}, \hat{C}_{O^{j}_{\alpha} V_{n} V_{n}}$ defined in (\ref{eq:3pfn}) are given for these operators by
\be
\label{eq:biopOPE1}
\hat{C}_{O^{j}_{\alpha} X_{n} X_{n}} =2C_{O_{\alpha}VV} C_{O_{\alpha}WW} +(n-2)C_{O_{\alpha}WW}^2, \quad \hat{C}_{O^{j}_{\alpha} V_{n} V_{n}}=n C_{O_{\alpha}VV}^2.
\ee
Similarly, the OPE coefficients involving the twist operators are given by \cite{Headrick:2010zt}\footnote{These OPE coefficients are obtained from two point functions on $\Sigma_n$ calculated via uniformization. We have adjusted the formula in \cite{Headrick:2010zt} to incorporate nonzero spin.} 
\be
\label{eq:biopOPE2}
\qquad  \hat{C}_{ O^{j}_{\alpha},\sigma_{n}  \tilde{\sigma}_{-n}}= \frac{1}{(-1)^{s_\alpha}}\f{1}{\left[2n\sin \left(\f{\pi j}{n}\right) \right]^{2\Delta}}.
\ee


\noindent 
We need to sum these up over $j$. By taking the Zamoldchikov metric into account, 
we get
\be
(-1)^{s_\alpha}\sum^{\f{n}{2}}_{j=1} C^{O^{j}_{\alpha} }_{\sigma_{n}  \tilde{\sigma}_{-n}}=\f{n}{2}\sum^{n-1}_{j=1} \f{1}{\left[2n\sin \left(\f{\pi j}{n}\right) \right]^{2\Delta}} \equiv \f{f(\Delta,n)}{2(2n)^{2\Delta}}.
\ee
where the above equation defines $f(\Delta,n)$.
We obtain $S_{n}(\rho_{V}|| \rho_{W})$ in the small subsystem limit, $1-\tilde{w}= 2\pi i x \ll 1$ by inserting these OPE coefficients in \eqref{eq:replicarelative} and summing up all the lightest primaries $\{O^{j}_{\alpha}\}$
\begin{align}
S_{n}(\rho_{V}|| \rho_{W})=\f{1}{n-1} \sum_{\alpha}\f{f(\Delta,n)}{2 n^{2\Delta}}\left(n C_{O_{\alpha}VV}^2-2C_{O_{\alpha}VV} C_{O_{\alpha}WW} -(n-2)C_{O_{\alpha}WW}^2 \right) (\pi x)^{2\Delta}.
\end{align}
Notice that the factor $(-1)^{s_\alpha}$ in \eqref{eq:biopOPE2} cancels with the identical factor coming from $(1-\tilde w)^{2h_\alpha}(1-\bar{ \tilde w})^{2\bar h_\alpha}$ in \eqref{eq:replicarelative}.
\noindent
To obtain the relative entropy we need to perform the analytic continuation to $n=1$. The sum
\be 
\f{f(\alpha,n)}{2} \equiv \sum^{n-1}_{l=1} \f{n-l}{\left(\sin \f{\pi l}{n}\right)^{2\alpha}} =  \sum^{n-1}_{m=1} \f{m}{\left(\sin \f{\pi m}{n}\right)^{2\alpha}} =\f{n}{2}\sum^{n-1}_{m=1} \f{1}{\left(\sin \f{\pi m}{n}\right)^{2\alpha}},
\label{eq:ansum}
\ee
can be analytically continued to $n \rightarrow 1$ by using the result of \cite{Calabrese:2010he}
\be 
\label{eq:fcont}
f(\alpha,n)= (n-1) \f{\Gamma (\f{3}{2})\Gamma (\alpha+1) }{\Gamma (\alpha+\f{3}{2})} +O((n-1)^2).
\ee 
Using this, we finally get the expression for the relative entropy
\be
S(\rho_{V}||\rho_{W})=\frac{\Gamma(\f{3}{2})\Gamma(\Delta+1)}{2\Gamma(\Delta+\f{3}{2})}\sum_{\alpha}\left(C_{O_{\alpha}VV}-C_{O_{\alpha}WW} \right)^2 (\pi x)^{2 \Delta}. \label{eq:relgen}
\ee

\noindent
If the lightest primaries satisfy $C_{O_{\alpha} VV}=C_{O_{\alpha} WW}$ then their descendants satisfy this as well. Therefore, we can actually take $\{ O_{\alpha} \}$ to be the set of the lightest primaries for which the above OPE differences do not vanish.

\noindent
Similarly one can pick up the leading term of the \eqref{eq:replicatracesquare} expansion for the trace square distance
\be
\label{eq:tracesqareequalweight}
T(\rho_{V}||\rho_{W})= \f{1}{2^{2\Delta}}\sum_{\alpha}\left(C_{O_{\alpha}VV}-C_{O_{\alpha}WW} \right)^2 (\pi x)^{2\Delta},
\ee

\noindent
therefore, they are essentially the same in the small interval limit

\be 
S(\rho_{V}||\rho_{W})=2^{2\Delta-1} \frac{\Gamma(\f{3}{2})\Gamma(\Delta+1)}{\Gamma(\Delta+\f{3}{2})} \;T(\rho_{V}||\rho_{W}).
\ee
Note that we have assumed $V$ and $W$ to be Hermitian operators in this section. When this is not the case, the formula clearly generalizes as
\be
S(\rho_{V}||\rho_{W})=\frac{\Gamma(\f{3}{2})\Gamma(\Delta+1)}{2\Gamma(\Delta+\f{3}{2})}\sum_{\alpha}\left(C_{O_{\alpha}V^*V}-C_{O_{\alpha}W^*W} \right)^2 (\pi x)^{2 \Delta}. \label{eq:relgenspin}
\ee
Here, the $\alpha$ sum runs over a Hermitian basis.


\subsection{Leading universal contribution when $h_V \neq h_W$}

Now we would like to return to the general expansions \eqref{eq:replicarelative} and \eqref{eq:replicatracesquare} and determine the leading behaviour in the subsystem size $x$. First, let us assume that there are no primaries in the original theory lighter than the stress tensor. In this case the leading contribution will come from some vacuum descendant states.

Let us first show that, similarly to the case of primary operators, single sheet insertions of vacuum descendants do not contribute to our expansions. Such a single sheet insertion has the form
\be
\mathcal{T}_{\{m_{i_{1}}\}}=[ L^{(1)}_{-\{m_{i_{1}}\}}\otimes I^{\otimes(n-1)} ]_{sym},
\ee
see \eqref{eq:virasoros} for the definition of $L^{(1)}_{-\{m_{i_{1}}\}}$.The subtlety compared to the case of primary operators is that the twist OPE coefficient $\hat{C}^{\mathcal{T}_{(k)}}_{\sigma_{n}\tilde{\sigma}_{-n}}$ can be nonvanishing. Still, because the OPE coefficients with the states $V_n$ and $X_n$ are
\bea
\hat{C}_{\mathcal{T}_{\{m_{i_{1}}\}}X_{n} X_{n}} =C_{L^{(1)}_{-\{m_{i_{1}}\}}VV} +(n-1) C_{L^{(1)}_{-\{m_{i_{1}}\}}WW}, && \hat{C}_{\mathcal{T}_{\{m_{i_{1}}\}}V_{n} V_{n}}=n C_{L^{(1)}_{-\{m_{i_{1}}\}}VV} ,
\eea
the combination $(\hat{C}_{V_{2} V_{2}\mathcal{T}_{\{m_{i_{1}}\}}}+\hat{C}_{W_{2} W_{2}\mathcal{T}_{\{m_{i_{1}}\}}}-2\hat{C}_{X_{2}X_{2}\mathcal{T}_{\{m_{i_{1}}\}}})$ in the expansion \eqref{eq:replicatracesquare} for the trace square distance automatically cancels. In the case of the relative entropy \eqref{eq:replicarelative} one has
\be
(\hat{C}_{X_{n}X_{n}\mathcal{T}_{\{m_{i_{1}}\}}}-\hat{C}_{V_{n} V_{n}\mathcal{T}_{\{m_{i_{1}}\}}})  = (n-1)(C_{L^{(1)}_{-\{m_{i_{1}}\}}VV}-C_{L^{(1)}_{-\{m_{i_{1}}\}}WW}),
\ee
so that the $\frac{1}{n-1}$ factor in front of $S_{n}( \rho_{V}|| \rho_{W})$ cancels. The OPE coefficient $\hat{C}^{\mathcal{T}_{\{m_{i_{1}}\}}}_{\sigma_{n}\tilde{\sigma}_{-n}}$ is proportional to the one point function $\langle L^{(1)}_{-\{m_{i_{1}}\}}\rangle_{\Sigma_n}$ which can then safely be taken to $n=1$ where it vanishes.

Now let us move on to discuss double sheet insertions, just like \eqref{eq:biop}, but now with $O_\alpha$ relpaced by some vacuum descendant. The lightest such operator comes from inserting two copies of the seed stress tensor $T$
\be
\label{eq:stressbiop}
\mathcal{T}^j = [T\otimes I^{\otimes(j-1)}\otimes T \otimes I^{\otimes(n-j-1)}]_{sym},\;\;\;j\leq \frac{n}{2}.
\ee
The OPE coefficients with the states $V_n$ and $X_n$ are the same as in \eqref{eq:biopOPE1} with $C_{VVT}=\langle T(\infty)V(1)V(0)\rangle=h_V$ and $C_{WWT}=h_W$. The twist OPE \eqref{eq:biopOPE2} is slightly modified by the one point function of the stress tensor on $\Sigma_n$:
\be
\label{eq:bistressope}
\qquad  \hat{C}_{ \mathcal{T}^j ,\sigma_{n}  \tilde{\sigma}_{-n}}= \f{\frac{c}{2}}{\left[2n\sin \left(\f{\pi j}{n}\right) \right]^{2\Delta}}+\langle T(0) \rangle_{\Sigma_n}^2,
\ee
where $\langle T(0) \rangle_{\Sigma_n}=\f{c}{24}\left(1-\frac{1}{n^2}\right)$ is the Schwartzian derivative of the uniformization map\footnote{Note that $\Sigma_n$ in this context has cuts between $u=1$ and $v=\infty$.} \eqref{eq:unifmap}.  As it turns out this has no effect on the $n\rightarrow 1$ continuation as clearly
\be 
\lim_{n\rightarrow 1} \frac{1}{2(n-1)}\sum_{j=1}^{n-1}\left[\f{c}{24}\left(1-\frac{1}{n^2}\right)\right]^2=0.
\ee
 We raise the index on $\hat{C}_{ \mathcal{T}^j ,\sigma_{n}  \tilde{\sigma}_{-n}}$ by using the Zamolodchikov metric
\bea
g^{\mathcal{T}^i\mathcal{T}^j}=\delta_{ij}n\left( \frac{2}{c}\right)^2, && j\leq \frac{n}{2}-1,
&& g^{\mathcal{T}^i\mathcal{T}^j}=\delta_{ij}\frac{n}{2}\left( \frac{2}{c}\right)^2, && j= \frac{n}{2}.
\eea
The analytic continuation can then be done in the same way as in the case of \eqref{eq:relgen} and the result is
\be
S(\rho_{V}||\rho_{W})=\f{16}{15}\f{1}{c}\left(h_V-h_W \right)^2 (\pi x)^{4} + \cdots, \label{eq:relT}
\ee
where we have inserted the scaling weight $\Delta=h+\bar h=2$ for the holomorphic stress tensor and an extra factor of 2 to take into account the identical contribution of the antiholomorphic stress tensor\footnote{Double sheet insertions containing a single holomorphic and a single antiholomorphic stress tensor vanish because the $\langle T \bar T\rangle$ two point function contains only contact terms.}. The universal contribution to the relative entropy between the vacuum and an excited state can be calculated in a different way, using the well known expression for the modular Hamiltonian of the vacuum. We use this to check \eqref{eq:relT} in appendix \ref{app:modh}.

We can calculate the contribution of the operators \eqref{eq:stressbiop} for the trace square distance in a similar manner. There is one key difference compared to the relative entropy: the Schwartzian derivative term in \eqref{eq:bistressope} does not drop out.  We find that
\be
T(\rho_{V}||\rho_{W})=\f{1}{4}\left(1+ \f{2}{c}\right)\left(h_V-h_W \right)^2 (\pi x)^{4} + \cdots. 
\ee
The $O(\frac{1}{c})$ term comes from the first term in \eqref{eq:bistressope} just like in the case of the relative entropy, while the $O(c^0)$ term comes from the contribution of the Schwartzian derivative in \eqref{eq:bistressope}. Notice that this latter term gives the contribution of the vacuum of the original theory, as can be seen by comparing this formula to \eqref{eq:tracedistsmalldistvac}. 

Finally, consider the case when there are primaries $O_{\alpha}$ in the spectrum with $\Delta<2$ and $C_{O_{\alpha}VV}-C_{O_{\alpha}WW}\neq 0$. In this case the leading small $x$ behaviour is given by \eqref{eq:relgen}, even when $h_V\neq h_W$, as both the lightest vacuum sector states and descendants of $O_{\alpha}$ contribute with higher powers of $x$. To see this result in action consider the example of a free scalar $X(z,\bar z)$. The relative entropy between vertex operators $\mathcal{V}_{\alpha}=e^{i\alpha X(z)}$ of weight $(h,\bar h)=\left(\frac{\alpha^2}{2},0 \right)$ 
was calculated in \cite{Lashkari:2015dia}
\be
\label{eq:freescalar}
S(\rho_{\mathcal{V}_\alpha}||\rho_{\mathcal{V}_\beta}) = (\alpha-\beta)^2(1-\pi x \cot (\pi x))=\frac{1}{3}(\alpha-\beta)^2(\pi x)^2 + O(x^4).
\ee
This is consistent with \eqref{eq:relgen} as we will now explain. We have two $U(1)$ currents $i\partial X(z)$  and $i\bar \partial X(\bar z)$ of dimension $\Delta=1$. The OPE coefficients are $C^{i\partial X}_{\mathcal{V}_{\alpha}\mathcal{V}_{-\alpha}}=-\alpha$ 
and $C^{ i\bar \partial X}_{\mathcal{V}_{\alpha}\mathcal{V}_{-\alpha}}=0$ 
\cite{francesco2012conformal}. The prefactor in \eqref{eq:relgen} for $\Delta=1$ is $\f{1}{3}$. We need to insert these into \eqref{eq:relgenspin}. Note that the current $i\partial X$ is Hermitian. 

\section{Generalized free fields} 
\label{sec:generalizedfreefields}

\subsection{Relative entropy}

In this section we discuss the relative entropy $S(\rho_{V}||\rho_{W}) $ of conformal field theories with a gravity dual. Generalized free fields $\{ \mathcal{O} \}$ are low energy excitations of such theories, whose correlation functions can be calculated by Wick contraction
\be 
\langle \mathcal{O} ^{*} (w_{0}) \cdots \mathcal{O} ^{*} (w_{n}) \mathcal{O}(\hat{w_{0}} ) \cdots \mathcal{O}(\hat{w_{n}} ) \rangle = \sum_{\sigma \in S_{n} } \prod_{j=0}^{n-1} \langle  \mathcal{O} ^{*} (w_{j})  \mathcal{O}  (\hat{w}_{\sigma(j)}) \rangle,  \label{eq: 2nptfn}
\ee
where $S_{n}$ denotes the symmetric group of order $n$ and $\sigma$ denotes an element of this group. 
\footnote{See Appendix \ref{app:genfree} for a  summary of the properties of the generalized free fields.}. In the dual gravity side these operators are identified with the minimally coupled bulk scalar fields.

Now we would like to compute the relative entropy between states created by two generalized free fields, $V,W$ with the same conformal dimension $h_{W}=h_{V}=h$, by using the replica trick (\ref{eq:relplica}).  These operators are located at
\be
w_{j}= e^{\f{2\pi ij}{n}}, \quad \hat{w}_{j}= e^{\f{2\pi i(j+x)}{n}}, 
\ee
 on the plane and each two point function appearing in (\ref{eq: 2nptfn}) is given by 
\be
\langle  \mathcal{O} ^{*} (w_{j})  \mathcal{O}  (\hat{w}_{\sigma(j)}) \rangle = \f{1}{\left(2\sin \f{\pi (j-\sigma(j)- x)}{n}\right)^{2\Delta_{\mathcal{O}}}}.
\ee
The $2n$ point function  (\ref{eq: 2nptfn}) does not contain the effects of Virasoro descendants. However, this does not cause any problems in the small subsystem limit $|A|=2\pi x \ll 1$ because of the following reason.  First of all, as we saw in the previous section, the  vacuum descendants do not contribute to the relative entropy. Furthermore, to get the leading result what we need to do is keeping only the first nontrivial primary exchange in the $2n$ point function, in this limit, so the inclusion of the stress energy exchanges in (\ref{eq: 2nptfn}) do not change the leading result.

To compute the relative entropy $S(\rho_{V}||\rho_{W}) $ with the use of the replica trick  (\ref{eq:relplica}), we need to perform the sum in (\ref{eq: 2nptfn}) explicitly, then analytically continue the result in $n$. In general they are both difficult tasks, however, we can perform them in the small interval limit $x \ll 1$. This is because in this limit, the dominant contribution in the sum over all elements of the symmetric group  $S_{n}$ in (\ref{eq: 2nptfn}) is coming from the identity, and the next to leading contributions are coming from the set of pair exchanges $\sigma_{a,b}$
\be
\sigma_{a,b}(a)=b, \quad \sigma_{a,b}(b)=a, \quad \sigma_{a,b}(k)=k,  \quad \forall k \neq a,b, \quad 0 \leq a,b,k \leq n-1.
\ee
In the small interval limit, we can neglect the contribution of the other elements of the group.  The $2n$ point function of the operator $V$ on the plane $C$ in this approximation is given by

\bea
\langle \prod^{n-1}_{k=0}& V^{*} (w_{k}) \cdots   \prod^{n-1}_{k=0}V (\hat{w}_{k})\rangle_{C} \\
&=\prod^{n-1}_{k=0} \langle V^{*} (w_{k}) V (\hat{w}_{k}) \rangle + \sum_{a,b=0, a\neq b}^{n-1}   \langle V^{*} (w_{a}) V (\hat{w}_{b}) \rangle  \langle V^{*} (w_{b}) V (\hat{w}_{a}) \rangle \prod_{k \neq a,b} \langle V^{*} (w_{k}) V (\hat{w}_{k}) \rangle \\
&=\f{1}{\left[2\sin \left( \f{\pi x}{n} \right) \right]^{4nh} } + \f{1}{\left[2\sin \left(\f{\pi x}{n} \right)\right]^{4(n-2) h} } \sum^{n-1}_{l=1} \f{(n-l)}{\left[2\sin \left(\f{\pi l}{n} \right)\right]^{8h}},
\eea

\vspace{0.5 cm}

\noindent Including the Jacobian factor (\ref{eq: Jacobian}), we obtain the $2n$ point function on the $n$ sheeted plane
\begin{align}
\langle \prod^{n-1}_{k=0}V^{\dagger} (\infty_{k})   \prod^{n-1}_{k=0}V (0_{k})\rangle_{\Sigma_{n}}&=\left[\f{2}{n}\sin \pi x \right]^{4nh}\langle \prod^{n-1}_{k=0}V^{*} (w_{k})  \prod^{n-1}_{k=0}V (\hat{w}_{k})\rangle \\
&=\left(\f{\sin \pi x}{\sin \f{\pi}{n} x}\right)^{4nh} \left[1+\f{f(4h ,n)}{2}\left(\sin \f{\pi x}{n}\right)^{8h} \right],
\end{align}
where $f(4h,n)$ is the same function as the one defined in (\ref{eq:ansum}). Similarly, we can compute the other $2n$ point function 
\bea
\langle & V^{\dagger} (\infty_{0}) \prod^{n-1}_{k=1}  W^{\dagger} (\infty_{k})    V(0_{0}) \prod^{n-1}_{k=1} W(0_{k})\rangle_{\Sigma_{n}} \\
 &=\left(\f{\sin \pi x}{\sin  \f{\pi x}{n}}  \right)^{4n h} \left[1 + \Big|\langle V | W \rangle\Big|^2\f{f(4h,n)}{n} \left(\sin  \f{\pi x}{n} \right)^{8h} \right. \left.+ \f{(n-2)f(4h,n)}{2n}\left(\sin  \f{\pi x}{n}  \right)^{8h} \right]. \label{eq:2nd2nptfn}
\eea
By using expressions (\ref{eq:2nd2nptfn}) and (\ref{eq: 2nptfn}) in \eqref{eq:4ptxn} along with the analytic continuation \eqref{eq:fcont} of $f$ we find the following formula for the relative entropy in the small interval limit $x \ll 1$, 
\be
S(\rho_{V}||\rho_{W}) =\f{\Gamma (\f{3}{2})\Gamma (4h+1) }{\Gamma (4h+\f{3}{2})} \left[1- \Big|\langle V | W \rangle\Big|^2 \right] (\pi x)^{8h}. \label{eq:relfree}
\ee
Note that the expression is symmetric under the exchange $ V \leftrightarrow W$.
It is satisfying that one can indeed reproduce (\ref{eq:relfree}) from the general formula (\ref{eq:relgenspin}) by using the OPE coefficients for the generalized free fields 
computed in Appendix \ref{app:genfree}. 
In particular, the first nontrivial primary appearing in the $V^{*}(z) \times V(0)$ OPE is just $(V^{*} V)(0)$, therefore $\Delta =2h_{V^{*} V}=4h$. This explains the $(\pi x)^{8h}$ behavior. Note that the state created by $V(0)$ is a single particle state of the bulk free field dual to $V$. Thus the holographic interpretation of \eqref{eq:relfree} is just the relative entropy between two single particle states of equal energy on a fixed background.

\subsection{Trace square distance}
Now we compute the exact trace square distance of generalized free fields. For these operators, the four point functions on the
plane are given by

\be
\langle V^{*} (w_{1}) V^{*} (w_{2})V (w_{3})V (w_{4})\rangle =\f{1}{|w_{13}w_{24}|^{4h}}+\f{1}{|w_{23}w_{14}|^{4h}} 
\ee
\be
\langle V^{*} (w_{1}) W^{*} (w_{2})V (w_{3})W (w_{4})\rangle =\f{1}{|w_{13}w_{24}|^{4h}}+\f{\Big|\langle V | W \rangle\Big|^2}{|w_{23}w_{14}|^{4h}}.
\ee
The positions of the insertions on the uniformized plane are given by
\be 
w_{1}=1, \quad w_{2}=-1, \quad  w_{3}=\hat{w}_{1}= e^{i\pi x} \quad , w_{4}=\hat{w}_{2}=-e^{i\pi x},
\ee
and using the Jacobian (\ref{eq: Jacobian})
we obtain the four point function on the two sheeted Riemann surface $\Sigma_{2}$
\be
\langle V^{*} (\infty_{1}) V^{*} (\infty_{2})V (0_{1})V (0_{2})\rangle_{\Sigma_{2}}=\left(\cos \f{\pi x}{2}\right)^{8h}+\left(\sin \f{\pi x}{2}\right)^{8h} ,
\ee
\be
\langle V^{*} (\infty_{1}) W^{*} (\infty_{2})V (0_{1})V (0_{2})\rangle_{\Sigma_{2}}=\left(\cos \f{\pi x}{2}\right)^{8h}+\Big|\langle V | W \rangle\Big|^2 \left(\sin \f{\pi x}{2}\right)^{8h} .
\ee
By using these and \eqref{eq:tsd} we get
\be
T(\rho_{V}||\rho_{W})=2  \left(\sin \f{\pi x}{2}\right)^{8h}  \left[1- \Big|\langle V | W \rangle\Big|^2 \right],
\ee
which can again be compared to \eqref{eq:tracesqareequalweight} using the formulae in Appendix \ref{app:genfree}.

\section{Critical Ising model}
\label{sec:ising}

Another excellent playground to test the validity of our formulae is the smallest minimal model, the critical Ising model in 2 dimensions. This theory posesses two distinct $(h,\bar h)=\left(\frac{1}{16},\frac{1}{16}\right)$ operators, the spin field $\sigma(z,\bar z)$ and the disorder operator $\mu(z,\bar z)$. Therefore, formula \eqref{eq:tracesqareequalweight} can be tested. They both fuse into the identity and the energy operator $\epsilon(z,\bar z)$ which has dimensions $\left(\frac{1}{2},\frac{1}{2}\right)$. According to \eqref{eq:tracesqareequalweight}, the trace square distance between $\sigma$ and $\mu$ is given by
\be
T(\rho_\sigma||\rho_\mu) = \frac{1}{4}(C_{\epsilon \sigma \sigma}-C_{\epsilon \mu \mu})^2 (\pi x)^2 + O(x^3).
\ee
The needed OPE coefficients are $C_{\epsilon \sigma \sigma}=\frac{1}{2}$ and $C_{\epsilon \mu \mu}=-\frac{1}{2}$ \cite{francesco2012conformal} and hence
\be
T(\rho_\sigma||\rho_\mu) = \frac{1}{4} (\pi x)^2 + O(x^3).
\ee
Now as all the $n$-point functions are known for minimal models, $T(\rho_\sigma||\rho_\mu)$ can be calculated \textit{exactly} using \eqref{eq:tsd} and \eqref{eq:uniformcorrelator}. The needed four point functions are\cite{francesco2012conformal}
\bea
\langle \sigma(z_1,\bar z_1)&\sigma(z_2,\bar z_2)\sigma(z_3,\bar z_3)\sigma(z_4,\bar z_4) \rangle \\
&=\left( \frac{1}{2} \frac{|z_{13}z_{24}|^{\frac{1}{2}}}{|z_{14}z_{23}z_{12}z_{34}|^{\frac{1}{2}}}\left[ 1+\frac{|z_{12}z_{34}|}{|z_{13}z_{24}|}+\frac{|z_{14}z_{23}|}{|z_{13}z_{24}|}\right]\right)^{\frac{1}{2}},\\
\langle \sigma(z_1,\bar z_1)&\mu(z_2,\bar z_2)\sigma(z_3,\bar z_3)\mu(z_4,\bar z_4) \rangle \\
&=\left( \frac{1}{2} \frac{|z_{13}z_{24}|^{\frac{1}{2}}}{|z_{14}z_{23}z_{12}z_{34}|^{\frac{1}{2}}}\left[ -1+\frac{|z_{12}z_{34}|}{|z_{13}z_{24}|}+\frac{|z_{14}z_{23}|}{|z_{13}z_{24}|}\right]\right)^{\frac{1}{2}},
\eea
where $z_{ij}=z_i-z_j$ and the $\langle \mu \mu \mu \mu \rangle$ four point function agrees with $\langle \sigma \sigma \sigma \sigma\rangle$. Some algebra reveals that
\be
T(\rho_\sigma||\rho_\mu) =2\left(1-\cos\frac{\pi x}{2}\right)=\frac{1}{4} (\pi x)^2 + O(x^4),
\ee
in accordance with our result.

\section{Conclusions} \label{sec:conc}

\vspace{0.1cm}

In this paper have we found the leading small interval behaviour of the relative entropy between two excited states with the same conformal dimension in 2d CFTs. 
We have also showed that in this limit the relative entropy is proportional to the trace square distance. The reason for this is that the correlation functions which are necessary to compute the
relative entropy in this limit are approximated by the same four point functions as the ones appearing in the trace square distance. We have checked 
our general results by computing the relative entropy between two generalized free fields by directly evaluating many point functions of these operators in the small interval limit. 

\vspace{0.1cm}

In addition, we have calculated the leading behaviour of the relative entropy when the conformal dimensions of the states are different. We have found that when there is a relevant primary in the OPE channel between the two excited states the leading behaviour of the relative entropy is dominated by this operator. Otherwise, the leading term is universal which is the case for CFTs with a gap. This is expected to be the case e.g. for theories describing pure gravity in three dimensions (given that they exist). For a general holographic theory we might have light excitations, for example bulk scalar fields. This modifies the leading behaviour of the relative entropy \cite{Blanco:2013joa} in accordance with the statement (\ref{point3}) of the introduction. 
\vspace{0.1cm}

Since formula \eqref{eq:relgen} is quite general, it would be nice if we could use it to learn more about some aspects of
the dynamics of black hole microstates. One of the key questions in recent debates about black holes is how much  can we  trust 
bulk effective field theories i.e. quantum field theories living on a fixed black hole background. One of the necessary condition for the validity of 
the bulk EFT is that the difference between black hole microstates is negligible. Therefore, the relative entropy provides a quantitative measure to check the validity of the bulk EFT or 
how much is it broken. According to \cite{Jafferis:2015del,Dong:2016eik}, our results should quantify distinguishability of bulk states with respect to measurments performed in the \textit{entanglement wedge} of region $A$. For static spacetimes, the deepest reach of this region into the bulk is given by the Ryu-Takayanagi surface anchored to $A$. In our case $A$ is small and this region is close to the boundary where we do not expect significant deviations from EFT\footnote{It seems reasonable to expect the OPE coefficients in \eqref{eq:relgen} to be of order $e^{-c}$ which is nonperturbative in $\frac{1}{c}$ as it should be. This is the case e.g. for three point functions in Liouville theory.}. In order to quantify the distinguishability of general states when the measurements are conducted close to the horizon one needs the relative entropy beyond our small subsystem limit. Indeed, the geodesic distance between the peak of the Ryu-Takayanagi surface and the closest point to it on the horizon in a static BTZ spacetime is given by
\be
\delta = \log \left( \coth \f{\pi r_h x}{2}\right) \approx e^{-\pi x r_h},
\ee
where $r_h$ is the location of the horizon in Schwartzschild coordinates and the geodesic reaches the boundary at angular coordinates $-\pi x$ and $\pi x$. The last approximation is valid when $r_h x$ is large. The horizon radius can be expressed with CFT data via $r_h=\sqrt{8G M} = 2 \sqrt{3 \Delta/c}$, where $\Delta=h+\bar h$. Therefore, we may probe a state $|V\rangle$ up to distances
\be
\delta \gtrsim e^{-2\pi x\sqrt{3 \Delta/c} }
\ee
from its event horizon. To keep this small when $x\rightarrow 0$, we need to consider states with $\frac{\Delta}{c} \sim \frac{1}{x^2}$. It is clear that we are loosing control of the approximation \eqref{eq:relgen} in this regime as the OPE coefficients implicitly depend on $x$ and this might change which intermediate state gives the leading behaviour. Equivalently, to probe EFT close to the horizon of a static black hole corresponding to a fixed value of $\Delta/c$, one needs the relative entropy accurately for interval sizes $2\pi x \gtrsim \sqrt{\frac{c}{3\Delta}}$.

%

\vspace{0.1cm}

It would also be interesting to quantify more precisely the distinguishability of two black hole microstates by computing the semiclassical limit of the OPE coefficients appearing in our formula. Notice that \eqref{eq:relgen} intrinsically contains details of the microscopics. Indeed, if we use some universal, classical limit for the OPE coefficients depending only on the operators weights and the central charge, \eqref{eq:relgen} gives zero automatically. Nevertheless, such a limit would be useful to obtain a more precise constraint on the magnitude of \eqref{eq:relgen} for large $c$.
In \cite{Chang:2015qfa}  it is argued from the semiclassical bootstrap analysis that for large $c$ CFTs with a sparse spectrum,
the square of the OPE coefficients summed over primary states with a fixed conformal dimension is universal and determined by the conformal blocks. 
Unfortunately, this universality is only confirmed when the conformal dimension of the internal operator is much larger than the dimensions of the external operators\footnote{ We thank Chi Ming Chang for discussions on this.}. 
It would be interesting to generalize this result and obtain a universal bound for the relative entropy between two black hole microstates.

\vspace{0.1cm}

Finally, note that there is a subtle connection with the eigenstate thermalization hypothesis (ETH) \cite{Sred, Deu,Garrison:2015lva}. According to ETH, at least when the relative size of the subsystem goes to zero, the reduced density matrix of any state should approach a thermal one $\rho_V\rightarrow \rho_{\beta_V}$ with some universal modular Hamiltonian. For 2d CFTs the temperature is related to the primary weight as $\frac{2\pi}{\beta_V}=\s{\f{24 h_{V}}{c}-1}$. It is easy to compute the relative entropy of two thermal states with $\beta_1 \approx \beta_2$. It reads as
\be
S(\rho_{\beta_1}||\rho_{\beta_2}) =-\frac{1}{\beta_1}\f{\p S(\rho_{\beta_1})}{\p \beta_1}(\beta_1-\beta_2)^2 + O((\beta_1-\beta_2)^3),
\ee
where $S$ is just the von Neumann entropy. For a 2d CFT, this formula predicts a $S(\rho_{\beta_1}||\rho_{\beta_2})\sim x^2$ start in the small subsystem limit which, according to the expansion \eqref{eq:replicarelative}, is only possible if there is a \textit{relevant} primary with scaling dimension $\Delta=1$ in the spectrum. This is the case e.g. for a free scalar, for which the relative entropy was computed in \cite{Lashkari:2015dia}. It would be interesting to learn more about the connection of this requirement to the nature of the limit $\rho_V\rightarrow \rho_{\beta_V}$.

\section*{Acknowledgments}
The authors thank Chi Ming Chang, Tarun Grover, Nima Lashkari, Andrea Phum, Joe Polchinski, Tadashi Takayanagi for discussions.
G. S. would like to thank the Gordon and Betty Moore Foundation for financial support. 
The work of G. S. and T. U. was supported in part by the National Science Foundation under Grant
No. NSF PHY-25915.

\appendix

\section{Relative entropy from the modular Hamiltonian}
\label{app:modh}

In this appendix we check the formula (\ref{eq:relT}) in the large central charge limit, when one of the excited states $|W \rangle$ is replaced by the ground state $|0 \rangle$, by combining some known results 
\cite{Blanco:2013joa,Wong:2013gua,Asplund:2014coa}. 
We start from the following expression of the relative entropy
\begin{align}
S(\rho_{V}|| \rho_{0})&= {\rm tr} \rho_{V} \log \rho_{V}- {\rm tr} \rho_{V} \log \rho_{0} \nonumber \\
&=\Delta \langle H_{{\rm vac}} \rangle -\Delta S,
\end{align}
where $\Delta S $ denotes the difference between the entanglement entropy of the excited state and the ground state and $H_{{\rm vac}} $ denotes the modular Hamiltonian of the vacuum reduced density matrix
\be 
H_{{\rm vac}}  \equiv - \log \rho_{0}, \quad  \Delta \langle H_{{\rm vac}} \rangle = {\rm tr} [\rho_{V}H_{{\rm vac}}] -
 {\rm tr} [\rho_{0}H_{{\rm vac}}].
\ee
In the large central charge limit with the ratio $h_{V}/c$ held fixed, the entanglement entropy of the excited state $|V \rangle $ is given by \cite{Asplund:2014coa}
\be 
S_{V}= \f{c}{3} \log \f{\beta_{V}}{\pi \epsilon} \sinh \f{\pi l}{\beta_{V}}, \quad \beta_{V} =\f{2\pi}{\s{\f{24h_{V}}{c}-1}}, \quad l=2 \pi x.
\ee
Combining this with the vacuum entanglement entropy
\be
S_{0} =\f{c}{3} \log \f{2}{\epsilon} \sin \f{l}{2},
\ee
in the small interval limit we get
\be
\Delta S= \f{h_{V}}{3} l^{2} +\left( \f{h_{V}}{180} -\f{h_{V}^2}{15c} \right) l^4 +O( l^6).
\ee
The Modular Hamiltonian of the cylinder vacuum is given by \cite{Blanco:2013joa}
\be
 H_{{\rm vac}} =2\pi \int^{l}_{0} d \phi \left[ \f{\cos (\phi-\f{l}{2})-\cos \f{l}{2}}{\sin \f{l}{2}} \right] T_{00},
\ee
where $T_{00}$ denotes the time  component of the stress energy tensor on the cylinder.
By using the relation
\be
2 \pi \langle V| T_{00}|V \rangle =\Delta_{V}=2h_{V},
\ee
we have
\be
\Delta \langle H_{{\rm vac}} \rangle =\f{h_{V}}{3} l^2 +\f{h_{V}}{180} l^4 + O(l^6).
\ee
Therefore, we get 
\be
S(\rho_{V}|| \rho_{0}) = \f{h_{V}^2}{15c} l^4 + O(l^6).
\ee
Since $l=2 \pi x$ we reproduce the result (\ref{eq:relT}).

\section{Correlation functions of generalized free fields}
\label{app:genfree}

In this appendix we calculate the correlation functions of generalized free fields.\footnote{For a review of generalized free fields, see \cite{kap}.} 
These operators are expanded as if they were free fields. For Hermitian fields the expansion is given by
\begin{align}
\label{eq:genfreefield}
V_{i} = \sum_{(n,\bar{n}) \in \mathbb{Z}^{+}} \f{1}{N_{(n,\bar{n})}}\left(\f{a^{V_{i}}_{n \bar{n}}}{z^{2h+n} \bar{z}^{2h+\bar{n}}} +z^{n} \bar{z}^{\bar{n}}a^{V_{i} \dagger}_{(n,\bar{n})} \right), \quad  0 \leq i \leq K
\end{align}
where $N_{n,\bar{n}}$ is a normalization factor,  $N_{(0,0)} =1$, and $h$ is the conformal dimension of $V_{i}$.
We can prescribe the following commutation relations between these creation and annihilation operators
\be
\left[ a^{V_{i}}_{n \bar{n}},\;a^{V_{j}}_{m \bar{m}} \right] = \left[ a^{V_{i} \dagger}_{n \bar{n}},\;a^{V_{j}\dagger}_{m \bar{m}} \right]=0, \quad  \left[ a^{V_{i} }_{n \bar{n}},\;a^{V_{j} \dagger}_{m \bar{m}} \right] =\delta^{ij} \delta_{nm}\delta_{\bar{n} \bar{m}}, 
\ee
and the definition of the vacuum
\be
a^{V_{i}}_{n \bar{n}} |0 \rangle =0,
\ee
which yield the following two point functions between the operators of \eqref{eq:genfreefield}
\be
\langle V_{i}(z,\bar{z}) V_{j}(0,0) \rangle = \f{\delta_{ij}}{(z\bar{z})^{2h}}.
\ee
In the calculation of the relative entropy, it is convenient to choose the complex basis,
\be
V=\f{V_{1}+iV_{2}}{\s{2}} \quad, V^{*}=\f{V_{1}-iV_{2}}{\s{2}},
\ee
so that the two point functions are
\be
\langle V^{*}(z,\bar{z}) V(0,0) \rangle = \f{1}{(z\bar{z})^{2h}}, \quad \langle V(z,\bar{z}) V(0,0) \rangle=\langle V^{*}(z,\bar{z}) V^{*}(0,0) \rangle=0. 
\ee
By taking some linear combination of $V_{i}$, one can introduce another generalized free field $W$
\be
W=\langle V|W \rangle V+ \langle V^{\perp}|W \rangle V^{\perp}, \quad \langle V|V^{\perp} \rangle=0.
\ee
 with the same conformal dimensions. The relevant two point functions are given by 
\be
\langle W^{*}(z) V(0) \rangle = \f{\langle W|V \rangle}{(z\bar{z})^{2h}}, \qquad \langle W(z) V(0) \rangle= \langle W^{*}(z) V^{*}(0) \rangle=0.
\ee
If we decompose this operator as 
\be
W=\f{W_{1}+iW_{2}}{\s{2}},\quad W^{*}=\f{W_{1}-iW_{2}}{\s{2}},
\ee
then the annihilation operators $\{ a^{V_{i}}_{n \bar{n}}, a^{W_{i}}_{n \bar{n}} \}, i=1,2$ satisfy
\be
\left[ a^{W_{1}}_{n \bar{n}}, a^{V_{1} \dagger}_{m \bar{m}} \right]=\left[ a^{W_{2}}_{n \bar{n}}, a^{V_{2} \dagger}_{m \bar{m}} \right]= \langle W| V\rangle \delta_{n \bar{n}} \delta_{m \bar{m}}, 
\quad \left[ a^{W_{1}}_{n \bar{n}}, a^{V^{\perp}_{1} \dagger}_{m \bar{m}} \right]=\left[ a^{W_{2}}_{n \bar{n}}, a^{V^{\perp}_{2} \dagger}_{m \bar{m}} \right]= \langle W| V^{\perp}\rangle \delta_{n \bar{n}} \delta_{m \bar{m}} \nonumber 
\ee
\be
\left[ a^{W_{2}}_{n \bar{n}}, a^{V_{1} \dagger}_{m \bar{m}} \right]=\left[ a^{W_{1}}_{n \bar{n}}, a^{V_{2} \dagger}_{m \bar{m}} \right]=\left[ a^{W_{2}}_{n \bar{n}}, a^{V^{\perp}_{1} \dagger}_{m \bar{m}} \right]=\left[ a^{W_{1}}_{n \bar{n}}, a^{V^{\perp}_{2} \dagger}_{m \bar{m}} \right]=0.
\ee
One can also infer the OPE structure of these generalized free fields. For example, by using the oscillator mode decomposition, we get
\be
V^{*}(z) \times V(0) \rightarrow \f{1}{(z\bar{z})^{2h}} +(V^{*}V)(0) +\cdots,
\ee
where $()$ denotes oscillator normal ordering. Therefore, 
\be
C_{V^{*}V (V^{*}V)}=1.
\ee
Similarly, one can compute the OPE coefficients involving $W$
\be
C_{W^{*}W(V^{*}V)}=|\langle V|W \rangle|^2, \quad C_{W^{*}W(V^{\perp*}V^{\perp})}=|\langle V^{\perp}|W \rangle|^2, \quad C_{W^{*}W(V^{*}V^{\perp})} = \langle V^{\perp}|W \rangle \langle W| V\rangle.
\ee
To calculate the small interval relative entropy from \eqref{eq:relgenspin} keep in mind that the intermediate operators in this formula must be Hermitian. The operator $(V^{*}V^{\perp})$ is not Hermitian, therefore we need to introduce linear combinations
\bea
O_1=\frac{(V^{*}V^{\perp})+(VV^{\perp *})}{\sqrt{2}}, && O_2=\frac{(V^{*}V^{\perp})-(VV^{\perp *})}{\sqrt{2}i},
\eea
for which the OPE coefficients are
\bea
C_{W^{*}W O_1}&=\frac{\langle V^{\perp}|W \rangle \langle W| V\rangle+\langle V|W \rangle \langle W| V^{\perp} \rangle}{\sqrt{2}}\\
C_{W^{*}W O_2}&=\frac{\langle V^{\perp}|W \rangle \langle W| V\rangle-\langle V|W \rangle \langle W| V^{\perp} \rangle}{\sqrt{2}i}
\eea
The dimension of these indermediate operators is $\Delta=4h$ so the relative entropy \eqref{eq:relgenspin} reads as
\bea
S(\rho_V||\rho_W) &= \frac{\Gamma(\f{3}{2})\Gamma(4h+1)}{2\Gamma(4h+\f{3}{2})} \\
&\times \left( (1-C_{W^{*}W(V^{*}V)})^2 + C_{W^{*}W(V^{\perp*}V^{\perp})}^2 + C_{W^{*}W O_1}^2+C_{W^{*}W O_2}^2 \right) (\pi x)^{8h}.
\eea
Now using that $\langle W|W\rangle=|\langle V|W\rangle|^2+|\langle V^{\perp}|W\rangle|^2=1$ we end up with
\be
S(\rho_V||\rho_W) = \frac{\Gamma(\f{3}{2})\Gamma(2h+1)}{2\Gamma(2h+\f{3}{2})} \left( 2-2|\langle V|W\rangle|^2\right) (\pi x)^{8h}.
\ee
Similarly, the trace square distance reads as
\be
T(\rho_V||\rho_W) = \frac{1}{2^{8h}}\left( 2-2|\langle V|W\rangle|^2\right) (\pi x)^{8h}.
\ee

{}


\newpage


\begin{thebibliography}{}


\bibitem{Maldacena:1997re} 
  J.~M.~Maldacena,
  ``The Large N limit of superconformal field theories and supergravity,''
  Int.\ J.\ Theor.\ Phys.\  {\bf 38}, 1113 (1999)
  [Adv.\ Theor.\ Math.\ Phys.\  {\bf 2}, 231 (1998)]
  doi:10.1023/A:1026654312961
  [hep-th/9711200].


\bibitem{Witten:1998qj} 
  E.~Witten,
  ``Anti-de Sitter space and holography,''
  Adv.\ Theor.\ Math.\ Phys.\  {\bf 2}, 253 (1998)
  [hep-th/9802150].


\bibitem{Gubser:1998bc} 
  S.~S.~Gubser, I.~R.~Klebanov and A.~M.~Polyakov,
  ``Gauge theory correlators from noncritical string theory,''
  Phys.\ Lett.\ B {\bf 428}, 105 (1998)
  doi:10.1016/S0370-2693(98)00377-3
  [hep-th/9802109].


\bibitem{Strominger:1996sh} 
  A.~Strominger and C.~Vafa,
  ``Microscopic origin of the Bekenstein-Hawking entropy,''
  Phys.\ Lett.\ B {\bf 379}, 99 (1996)
  doi:10.1016/0370-2693(96)00345-0
  [hep-th/9601029].


\bibitem{Callan:1996dv} 
  C.~G.~Callan and J.~M.~Maldacena,
  ``D-brane approach to black hole quantum mechanics,''
  Nucl.\ Phys.\ B {\bf 472}, 591 (1996)
  doi:10.1016/0550-3213(96)00225-8
  [hep-th/9602043].

\bibitem{Strominger:1997eq} 
  A.~Strominger,
  ``Black hole entropy from near horizon microstates,''
  JHEP {\bf 9802}, 009 (1998)
  doi:10.1088/1126-6708/1998/02/009
  [hep-th/9712251].


\bibitem{Balasubramanian:2005qu} 
  V.~Balasubramanian, P.~Kraus and M.~Shigemori,
  ``Massless black holes and black rings as effective geometries of the D1-D5 system,''
  Class.\ Quant.\ Grav.\  {\bf 22}, 4803 (2005)
  doi:10.1088/0264-9381/22/22/010
  [hep-th/0508110].

\bibitem{Balasubramanian:2005mg} 
  V.~Balasubramanian, J.~de Boer, V.~Jejjala and J.~Simon,
  ``The Library of Babel: On the origin of gravitational thermodynamics,''
  JHEP {\bf 0512}, 006 (2005)
  doi:10.1088/1126-6708/2005/12/006
  [hep-th/0508023].

\bibitem{Balasubramanian:2007qv} 
  V.~Balasubramanian, B.~Czech, V.~E.~Hubeny, K.~Larjo, M.~Rangamani and J.~Simon,
  ``Typicality versus thermality: An Analytic distinction,''
  Gen.\ Rel.\ Grav.\  {\bf 40}, 1863 (2008)
  doi:10.1007/s10714-008-0606-8
  [hep-th/0701122].

\bibitem{Das:2008ka} 
  S.~R.~Das and G.~Mandal,
  ``Microstate Dependence of Scattering from the D1-D5 System,''
  JHEP {\bf 0904}, 036 (2009)
  doi:10.1088/1126-6708/2009/04/036
  [arXiv:0812.1358 [hep-th]].


\bibitem{Hawking:1976ra} 
  S.~W.~Hawking,
  ``Breakdown of Predictability in Gravitational Collapse,''
  Phys.\ Rev.\ D {\bf 14}, 2460 (1976).
  doi:10.1103/PhysRevD.14.2460



\bibitem{Hawking:1974sw} 
  S.~W.~Hawking,
  ``Particle Creation by Black Holes,''
  Commun.\ Math.\ Phys.\  {\bf 43}, 199 (1975)
  [Commun.\ Math.\ Phys.\  {\bf 46}, 206 (1976)].
  doi:10.1007/BF02345020


\bibitem{Mathur:2005zp} 
  S.~D.~Mathur,
  ``The Fuzzball proposal for black holes: An Elementary review,''
  Fortsch.\ Phys.\  {\bf 53}, 793 (2005)
  doi:10.1002/prop.200410203
  [hep-th/0502050].





\bibitem{Casini:2008cr} 
  H.~Casini,
  ``Relative entropy and the Bekenstein bound,''
  Class.\ Quant.\ Grav.\  {\bf 25}, 205021 (2008)
  doi:10.1088/0264-9381/25/20/205021
  [arXiv:0804.2182 [hep-th]].

\bibitem{Wall:2010cj} 
  A.~C.~Wall,
  ``A Proof of the generalized second law for rapidly-evolving Rindler horizons,''
  Phys.\ Rev.\ D {\bf 82}, 124019 (2010)
  doi:10.1103/PhysRevD.82.124019
  [arXiv:1007.1493 [gr-qc]].

\bibitem{Wall:2011hj} 
  A.~C.~Wall,
  ``A proof of the generalized second law for rapidly changing fields and arbitrary horizon slices,''
  Phys.\ Rev.\ D {\bf 85}, 104049 (2012)
  [Phys.\ Rev.\ D {\bf 87}, no. 6, 069904 (2013)]
  doi:10.1103/PhysRevD.87.069904, 10.1103/PhysRevD.85.104049
  [arXiv:1105.3445 [gr-qc]].

\bibitem{Bousso:2014sda} 
  R.~Bousso, H.~Casini, Z.~Fisher and J.~Maldacena,
  ``Proof of a Quantum Bousso Bound,''
  Phys.\ Rev.\ D {\bf 90}, no. 4, 044002 (2014)
  doi:10.1103/PhysRevD.90.044002
  [arXiv:1404.5635 [hep-th]].


\bibitem{Bousso:2014uxa} 
  R.~Bousso, H.~Casini, Z.~Fisher and J.~Maldacena,
  ``Entropy on a null surface for interacting quantum field theories and the Bousso bound,''
  Phys.\ Rev.\ D {\bf 91}, no. 8, 084030 (2015)
  doi:10.1103/PhysRevD.91.084030
  [arXiv:1406.4545 [hep-th]].

\bibitem{Lin:2014hva} 
  J.~Lin, M.~Marcolli, H.~Ooguri and B.~Stoica,
  ``Locality of Gravitational Systems from Entanglement of Conformal Field Theories,''
  Phys.\ Rev.\ Lett.\  {\bf 114}, 221601 (2015)
  doi:10.1103/PhysRevLett.114.221601
  [arXiv:1412.1879 [hep-th]].

\bibitem{Lashkari:2014kda} 
  N.~Lashkari, C.~Rabideau, P.~Sabella-Garnier and M.~Van Raamsdonk,
  ``Inviolable energy conditions from entanglement inequalities,''
  JHEP {\bf 1506}, 067 (2015)
  doi:10.1007/JHEP06(2015)067
  [arXiv:1412.3514 [hep-th]].

\bibitem{Lashkari:2015hha} 
  N.~Lashkari and M.~Van Raamsdonk,
  ``Canonical Energy is Quantum Fisher Information,''
  arXiv:1508.00897 [hep-th].


\bibitem{Blanco:2013joa} 
  D.~D.~Blanco, H.~Casini, L.~Y.~Hung and R.~C.~Myers,
  ``Relative Entropy and Holography,''
  JHEP {\bf 1308}, 060 (2013)
  doi:10.1007/JHEP08(2013)060
  [arXiv:1305.3182 [hep-th]].
  


\bibitem{Lashkari:2013koa} 
  N.~Lashkari, M.~B.~McDermott and M.~Van Raamsdonk,
  ``Gravitational dynamics from entanglement 'thermodynamics',''
  JHEP {\bf 1404}, 195 (2014)
  doi:10.1007/JHEP04(2014)195
  [arXiv:1308.3716 [hep-th]].


\bibitem{Faulkner:2013ica}
  T.~Faulkner, M.~Guica, T.~Hartman, R.~C.~Myers and M.~Van Raamsdonk,
  ``Gravitation from Entanglement in Holographic CFTs,''
  JHEP {\bf 1403} (2014) 051
  doi:10.1007/JHEP03(2014)051
  [arXiv:1312.7856 [hep-th]].



\bibitem{Jafferis:2015del} 
  D.~L.~Jafferis, A.~Lewkowycz, J.~Maldacena and S.~J.~Suh,
  ``Relative entropy equals bulk relative entropy,''
  arXiv:1512.06431 [hep-th].

\bibitem{Bhattacharya:2012mi} 
  J.~Bhattacharya, M.~Nozaki, T.~Takayanagi and T.~Ugajin,
  ``Thermodynamical Property of Entanglement Entropy for Excited States,''
  Phys.\ Rev.\ Lett.\  {\bf 110}, no. 9, 091602 (2013)
  doi:10.1103/PhysRevLett.110.091602
  [arXiv:1212.1164].



\bibitem{Lashkari:2015dia} 
  N.~Lashkari,
  ``Modular Hamiltonian of Excited States in Conformal Field Theory,''
  arXiv:1508.03506 [hep-th].

\bibitem{Lashkari:2014yva} 
  N.~Lashkari,
  ``Relative Entropies in Conformal Field Theory,''
  Phys.\ Rev.\ Lett.\  {\bf 113}, 051602 (2014)
  doi:10.1103/PhysRevLett.113.051602
  [arXiv:1404.3216 [hep-th]].

\bibitem{Alcaraz:2011tn} 
  F.~C.~Alcaraz, M.~I.~Berganza and G.~Sierra,
  ``Entanglement of low-energy excitations in Conformal Field Theory,''
  Phys.\ Rev.\ Lett.\  {\bf 106}, 201601 (2011)
  doi:10.1103/PhysRevLett.106.201601
  [arXiv:1101.2881 [cond-mat.stat-mech]].


\bibitem{Caputa:2014vaa} 
  P.~Caputa, M.~Nozaki and T.~Takayanagi,
  ``Entanglement of local operators in large-N conformal field theories,''
  PTEP {\bf 2014}, 093B06 (2014)
  [arXiv:1405.5946 [hep-th]].

\bibitem{Asplund:2014coa} 
  C.~T.~Asplund, A.~Bernamonti, F.~Galli and T.~Hartman,
  ``Holographic Entanglement Entropy from 2d CFT: Heavy States and Local Quenches,''
  JHEP {\bf 1502}, 171 (2015)
  [arXiv:1410.1392 [hep-th]].


\bibitem{Holzhey:1994we} 
  C.~Holzhey, F.~Larsen and F.~Wilczek,
  ``Geometric and renormalized entropy in conformal field theory,''
  Nucl.\ Phys.\ B {\bf 424}, 443 (1994)
  doi:10.1016/0550-3213(94)90402-2
  [hep-th/9403108].

\bibitem{Calabrese:2004eu} 
  P.~Calabrese and J.~L.~Cardy,
  ``Entanglement entropy and quantum field theory,''
  J.\ Stat.\ Mech.\  {\bf 0406}, P06002 (2004)
  doi:10.1088/1742-5468/2004/06/P06002
  [hep-th/0405152].


\bibitem{Cardy:2007mb} 
  J.~L.~Cardy, O.~A.~Castro-Alvaredo and B.~Doyon,
  ``Form factors of branch-point twist fields in quantum integrable models and entanglement entropy,''
  J.\ Statist.\ Phys.\  {\bf 130}, 129 (2008)
  doi:10.1007/s10955-007-9422-x
  [arXiv:0706.3384 [hep-th]].
  
  
\bibitem{Calabrese:2009qy} 
  P.~Calabrese and J.~Cardy,
  ``Entanglement entropy and conformal field theory,''
  J.\ Phys.\ A {\bf 42}, 504005 (2009)
  doi:10.1088/1751-8113/42/50/504005
  [arXiv:0905.4013 [cond-mat.stat-mech]].
  
  
\bibitem{Berganza:2011mh} 
  M.~I.~Berganza, F.~C.~Alcaraz and G.~Sierra,
  ``Entanglement of excited states in critical spin chians,''
  J.\ Stat.\ Mech.\  {\bf 1201}, P01016 (2012)
  doi:10.1088/1742-5468/2012/01/P01016
  [arXiv:1109.5673 [cond-mat.stat-mech]].


  
\bibitem{Calabrese:2012ew} 
  P.~Calabrese, J.~Cardy and E.~Tonni,
  ``Entanglement negativity in quantum field theory,''
  Phys.\ Rev.\ Lett.\  {\bf 109}, 130502 (2012)
  doi:10.1103/PhysRevLett.109.130502
  [arXiv:1206.3092 [cond-mat.stat-mech]].

\bibitem{Calabrese:2014yza} 
  P.~Calabrese, J.~Cardy and E.~Tonni,
  ``Finite temperature entanglement negativity in conformal field theory,''
  J.\ Phys.\ A {\bf 48}, no. 1, 015006 (2015)
  doi:10.1088/1751-8113/48/1/015006
  [arXiv:1408.3043 [cond-mat.stat-mech]].



\bibitem{Nozaki:2014uaa} 
  M.~Nozaki,
  ``Notes on Quantum Entanglement of Local Operators,''
  JHEP {\bf 1410}, 147 (2014)
  doi:10.1007/JHEP10(2014)147
  [arXiv:1405.5875 [hep-th]].



\bibitem{Caputa:2014eta} 
  P.~Caputa, J.~Simón, A.~Štikonas and T.~Takayanagi,
  ``Quantum Entanglement of Localized Excited States at Finite Temperature,''
  JHEP {\bf 1501}, 102 (2015)
  doi:10.1007/JHEP01(2015)102
  [arXiv:1410.2287 [hep-th]].


\bibitem{Caputa:2015waa} 
  P.~Caputa, J.~Simón, A.~Štikonas, T.~Takayanagi and K.~Watanabe,
  ``Scrambling time from local perturbations of the eternal BTZ black hole,''
  JHEP {\bf 1508}, 011 (2015)
  doi:10.1007/JHEP08(2015)011
  [arXiv:1503.08161 [hep-th]].

\bibitem{Mosaffa:2012mz} 
  A.~E.~Mosaffa,
  ``Symmetric Orbifolds and Entanglement Entropy for Primary Excitations in Two Dimensional CFT,''
  arXiv:1208.3204 [hep-th].

\bibitem{Giusto:2014aba} 
  S.~Giusto and R.~Russo,
  ``Entanglement Entropy and D1-D5 geometries,''
  Phys.\ Rev.\ D {\bf 90}, no. 6, 066004 (2014)
  doi:10.1103/PhysRevD.90.066004
  [arXiv:1405.6185 [hep-th]].

\bibitem{NC}
M. ~Nielsen, I.~ Chuang,
 ``Quantum Computation and Quantum information,''
Cambridge University Press

\bibitem{Calabrese:2009ez} 
  P.~Calabrese, J.~Cardy and E.~Tonni,
  ``Entanglement entropy of two disjoint intervals in conformal field theory,''
  J.\ Stat.\ Mech.\  {\bf 0911}, P11001 (2009)
  doi:10.1088/1742-5468/2009/11/P11001
  [arXiv:0905.2069 [hep-th]].

\bibitem{Calabrese:2010he} 
  P.~Calabrese, J.~Cardy and E.~Tonni,
  ``Entanglement entropy of two disjoint intervals in conformal field theory II,''
  J.\ Stat.\ Mech.\  {\bf 1101}, P01021 (2011)
  doi:10.1088/1742-5468/2011/01/P01021
  [arXiv:1011.5482 [hep-th]].


%
%
%


\bibitem{Chang:2015qfa} 
  C.~M.~Chang and Y.~H.~Lin,
  ``Bootstrapping 2D CFTs in the Semiclassical Limit,''
  arXiv:1510.02464 [hep-th].


\bibitem{Headrick:2010zt} 
  M.~Headrick,
  ``Entanglement Renyi entropies in holographic theories,''
  Phys.\ Rev.\ D {\bf 82}, 126010 (2010)
  doi:10.1103/PhysRevD.82.126010
  [arXiv:1006.0047 [hep-th]].
  

\bibitem{kap}
J.~Kaplan
``Lectures on AdS/CFT from the Bottom Up"
http://www.pha.jhu.edu/~jaredk/AdSCFTCourseNotesPublic.pdf

\bibitem{francesco2012conformal}
 P. ~Francesco P. ~Mathieu and D. ~S{\'e}n{\'e}chal,
 ``Conformal field theory,''
  2012,
 Springer Science \& Business Media
 
\bibitem{Dong:2016eik} 
  X.~Dong, D.~Harlow and A.~C.~Wall,
  ``Bulk Reconstruction in the Entanglement Wedge in AdS/CFT,''
  arXiv:1601.05416 [hep-th].
  

%
%


\bibitem{Sred}
M, Srednicki, 
``Chaos and quantum thermalization''	
Physical Review E (Statistical Physics, Plasmas, Fluids, and Related Interdisciplinary Topics), Volume 50, Issue 2, August 1994, pp.888-901




\bibitem{Deu}
  J.M. Deutsch, 
(February 1991). "Quantum statistical mechanics in a closed system". Physical Review A 43 (4): 2046-2049

\bibitem{Garrison:2015lva} 
  J.~R.~Garrison and T.~Grover,
  ``Does a single eigenstate encode the full Hamiltonian?,''
  arXiv:1503.00729 [cond-mat.str-el].
  
\bibitem{Wong:2013gua}
  G.~Wong, I.~Klich, L.~A.~Pando Zayas and D.~Vaman,
  ``Entanglement Temperature and Entanglement Entropy of Excited States,''
  JHEP {\bf 1312} (2013) 020
  [arXiv:1305.3291 [hep-th]].


\end{thebibliography}
\end{document}